\documentclass[aps,
reprint,
superscriptaddress,
amsmath, amssymb,
floatfix
]{revtex4-2}

\usepackage{graphicx}
\usepackage{subfigure}
\usepackage{multirow}
\usepackage{dcolumn}
\usepackage{bm}
\usepackage{amsfonts}
\usepackage{mathrsfs}
\usepackage{color}

\usepackage{hyperref}
\usepackage{cleveref}
\crefname{figure}{\textcolor{black}{Fig.}}{\textcolor{black}{Fig.}}
\crefname{table}{\textcolor{black}{Tab.}}{\textcolor{black}{Tab.}}
\crefname{equation}{\textcolor{black}{Eq.}}{\textcolor{black}{Eq.}}
\crefname{section}{\textcolor{black}{Sec.}}{\textcolor{black}{Sec.}}

\begin{document}
\preprint{APS/123-QED}

\title{Quasiperiodic {$[110]$} Symmetric Tilt FCC Grain Boundaries}
\author{Wenwen Zou}
\affiliation{Hunan Key Laboratory for Computation and Simulation in Science and Engineering, Key Laboratory of Intelligent Computing and Information Processing of Ministry of Education, School of Mathematics and Computational Science, Xiangtan University, Xiangtan, Hunan, China, 411105}

\author{Juan Zhang}
\affiliation{Hunan Key Laboratory for Computation and Simulation in Science and Engineering, Key Laboratory of Intelligent Computing and Information Processing of Ministry of Education, School of Mathematics and Computational Science, Xiangtan University, Xiangtan, Hunan, China, 411105}

\author{Jie Xu}
\email{xujie@lsec.cc.ac.cn}
\affiliation{LSEC \& NCMIS, Institute of Computational Mathematics and Scientific/Engineering Computing, Academy of Mathematics and Systems Science, Chinese Academy of Sciences, Beijing, China}

\author{Kai Jiang}
\email{kaijiang@xtu.edu.cn}
\affiliation{Hunan Key Laboratory for Computation and Simulation in Science and Engineering, Key Laboratory of Intelligent Computing and Information Processing of Ministry of Education, School of Mathematics and Computational Science, Xiangtan University, Xiangtan, Hunan, China, 411105}

\date{\today}

\begin{abstract}
In this work, we investigate {$\left[ 110 \right]$} symmetric tilt FCC grain boundaries (GBs) by a recently developed approach for quasiperiodic interfaces using the Landau-Brazovskii model.
On special tilt angles associated with quadratic algebraic numbers, quasiperiodic GBs exhibit generalized Fibonacci substitution rules.
The transition mechanism from quasiperiodic GBs to periodic GBs is explored through sphere offsets and spectral coalescence.
We also propose an accurate method to calculate the GB energy for arbitrary tilt angle and analyze the factors affecting the GB energy.
The revealing GB energy change is continue along with tilt angle except periodic GBs.
\end{abstract}

\maketitle

Face-centered cubic (FCC) structures are observed in a wide range of scales from atomic/ionic crystals\,\cite {yablonovitch1991photonic} to supramolecular colloids\,\cite{pileni1998colloidal} and block copolymers\,\cite{huang2003face}. 
The properties of FCC materials, including plastic deformation\,\cite{jiang2023stabilizing}, strength\,\cite{guan2020possibility}, fatigue\,\cite{guan2022pathway} and creep\,\cite{li2023twin}, are greatly influenced by grain boundaries (GBs). 
Experimental studies on FCC GBs concentrate on specific orientations\,\cite{randle1991influence,guan2022pathway,guan2020possibility} that exhibit higher symmetry, such as symmetric tilt GBs along the close-packed direction $\left[ 110 \right]$ (Miller indices)\,\cite{tschopp2015symmetric}. 
In these particular GBs, two grains could generate a coincident site lattice (CSL)\,\cite{friedel1926leccons,straumal2005faceting}, whose index $\Sigma$ measures the size of the common unit cell of two grains. 
Theoretical studies, whether atom-based\,\cite{rittner1996110,liu1998grain,zheng2020grain,tschopp2015symmetric} or field-based approaches\,\cite{bulatov2014grain,su2021phasefield,blixt2022evaluation}, mainly focus on low-$\Sigma$ CSL GBs under periodic boundary conditions. 

GBs of general orientations are usually non-CSL\,\cite{li2019atomistic}, in which nonperiodic ordered sequences or patterns have been observed experimentally\,\cite{guyoncourt1968deformation,ross1969type}.
Non-CSL GBs have been conceptually viewed as quasiperiodic structures because they intuitively stem from irrational 2D slices of 3D periodic structure\,\cite{sutton1981structure,bollmann1982crystal,sutton1983structure,gratias1988hidden,sutton1995interfaces}. 
The quasiperiodicity is implied by finite-size simulations\,\cite{van2001grain,lee2004computation,li2019atomistic}, where partial structures can be captured but the long-range configurations have been not yet revealed. 

In this work, we investigate the general $\left[ 110 \right]$ symmetric tilt FCC GBs, especially the quasiperiodic GBs.
We adopt the Landau-Brazovskii (LB) model\,\cite{brazovskii1975phase,fredrickson1987fluctuation} and the recently developed approach to deal with GBs and interfaces of general orientations\,\cite{xu2017computing,cao2021computing,jiang2022tilt,chen2024investigation,jiang2014numerical,jiang2018numerical,jiang2024numerical,GBsoftware}. 
Two quasiperiodic GBs exhibiting generalized Fibonacci sequences are discovered, and the transition mechanism from quasiperiodic GBs to periodic GBs is investigated from the viewpoints of structure and spectra.
An accurate method to calculate GB energy is proposed and GB widths are evaluated for tilt angles of both periodic and quasiperiodic cases.

The LB energy per volume of a scalar field $\phi(\boldsymbol{r})$ is\,\cite{brazovskii1975phase}
\begin{equation}\label{eq:LBmodel}
\begin{aligned}
	E[\phi]= \frac{1}{V} & \int_{\Omega} \left\lbrace \frac{1}{2}\left[(\Delta+1) \phi\right]^{2} 
    +\frac{\tau}{2!} \phi^{2}-\frac{\gamma}{3!} \phi^{3}+\frac{1}{4!} \phi^{4} \right\rbrace \mathrm{d} \boldsymbol{r}
\end{aligned}
\end{equation}
in the region $\Omega$ of the volume $V$.
The coefficients, while can be related to physical parameters\,\cite{fredrickson1987fluctuation,leibler1980theory}, are chosen as $\tau = 0.2$ and $\gamma = 1.5$ in the FCC region of the phase diagram\,\cite{mcclenagan2019landau}. 
The bulk FCC profile is obtained by minimizing the LB free energy in a cubic unit cell using periodic Fourier expansion, meanwhile optimizing the cell size $a$\,\cite{jiang2020efficient,bao2024convergence}.

\begin{figure*}
	\centering
	\subfigure[$\Sigma 3\,(111) $ GB]{
		\includegraphics[width=0.3\textwidth]{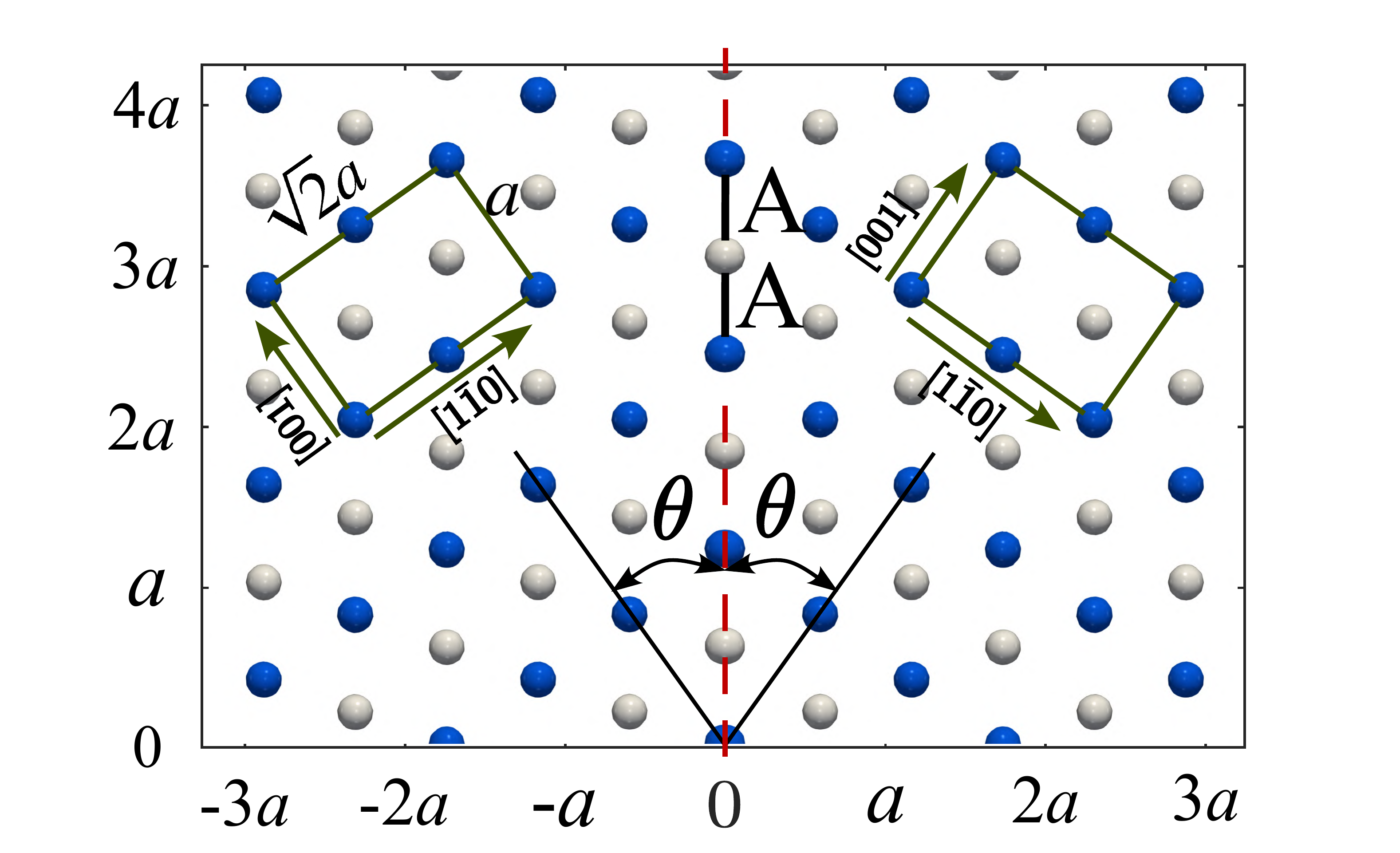}}
	\subfigure[$\Sigma 11\,(113) $ GB]{
		\includegraphics[width=0.3\textwidth]{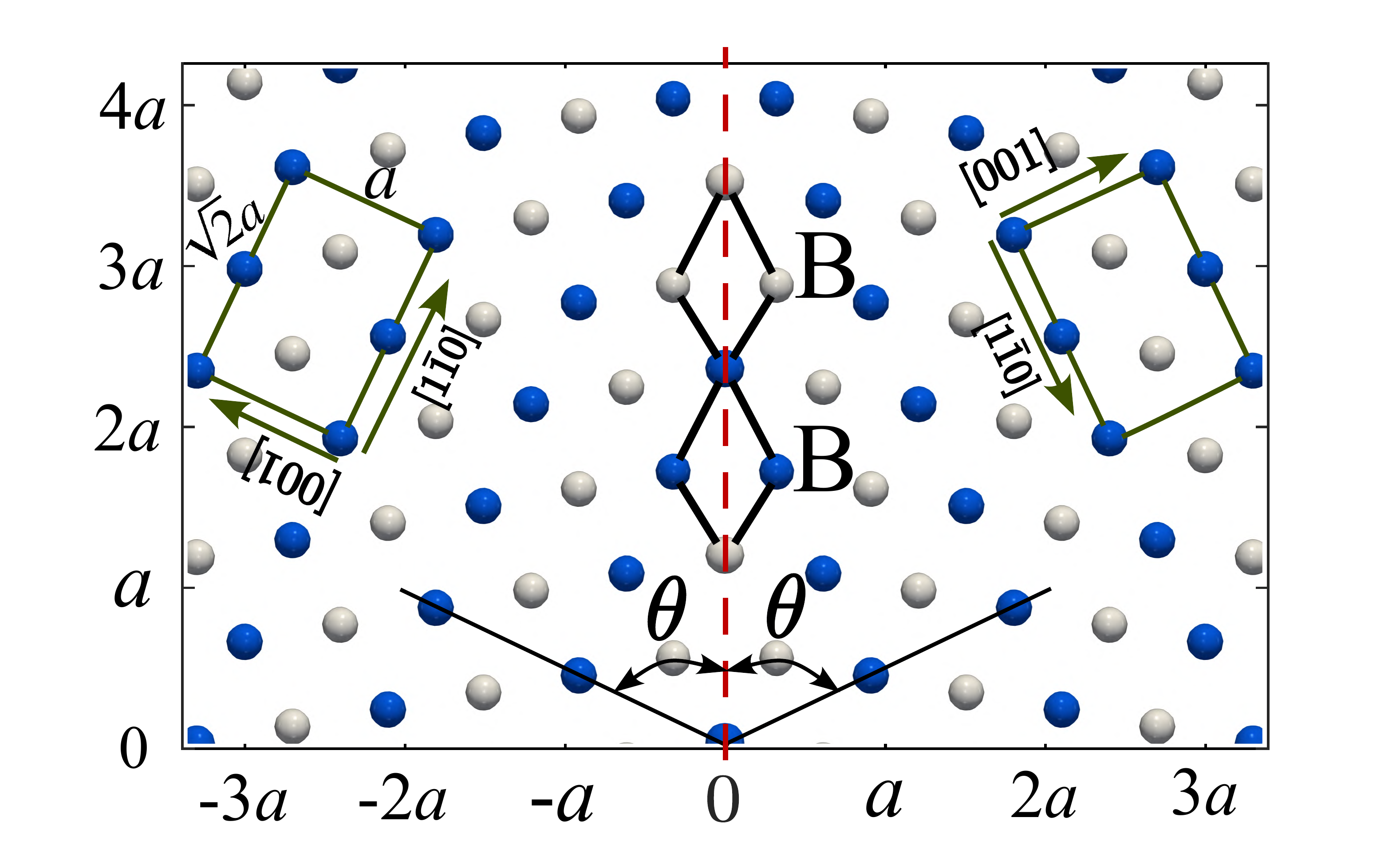}}
	\subfigure[$\Sigma 3\,(112) $ GB]{
		\includegraphics[width=0.3\textwidth]{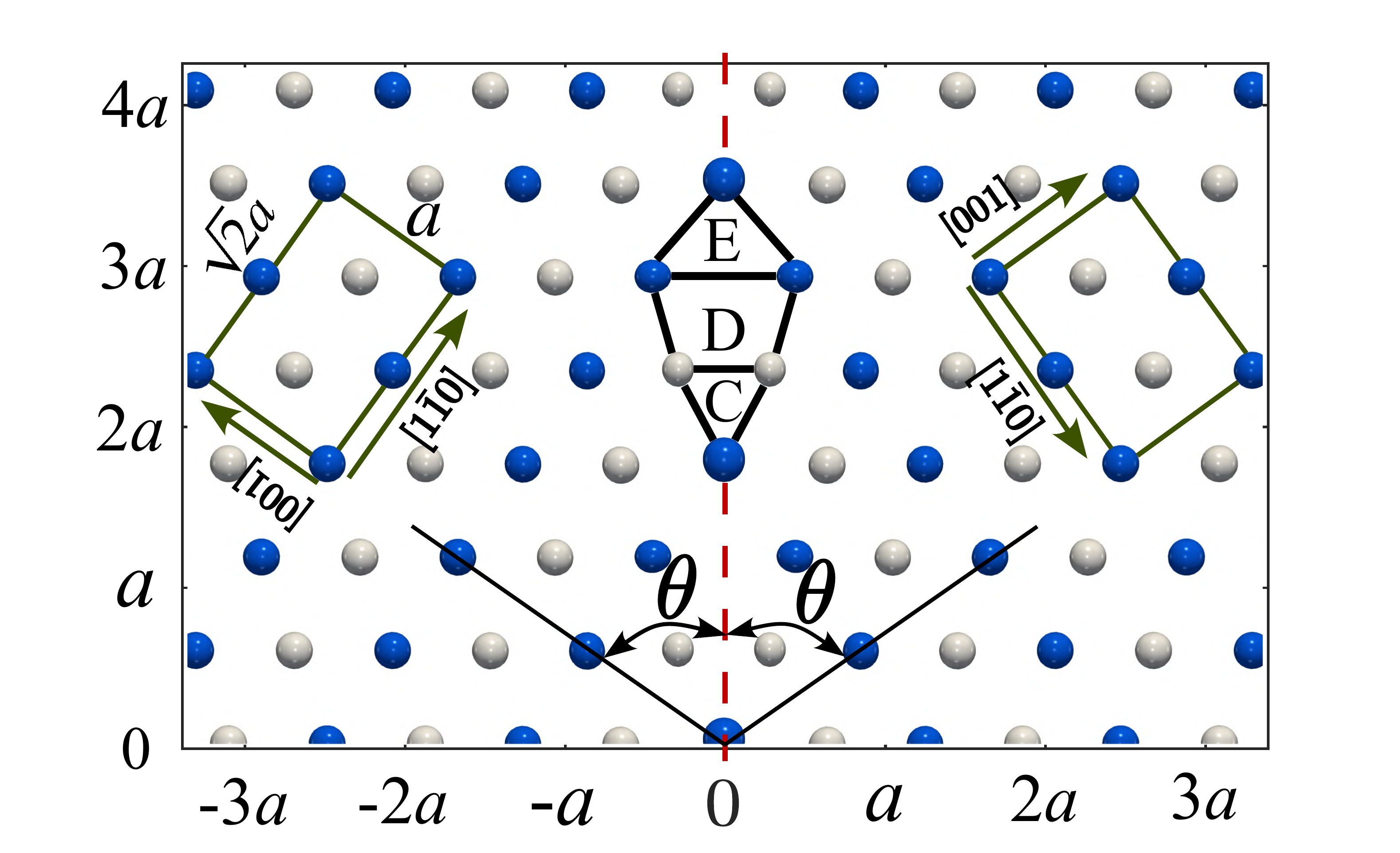}}
	\caption{Periodic GBs: (a) $\Sigma 3\,(111) $ GB; (b) $\Sigma 11\,(113) $ GB; (c) $\Sigma 3\,(112) $ GB.
	The red dashed line indicates the $x=0$ plane.
    Structural units on GBs are marked with solid black lines.
    Tilt angles, single cells and grain orientations are labeled.}
	\label{fig:periodic_GBs}
\end{figure*}

To setup the computation of GBs, two FCC grains with prescribed orientations are placed in two halfspaces, $x<0$ (grain 1) and $x>0$ (grain 2). 
The $\left[ 110 \right]$ directions of two grains coincide and align with the $y$-axis, while the $\left[ 001 \right]$ direction of grain 1 (resp. grain 2) is rotated to $(-\sin\theta,0,\cos\theta)$ (resp. $(\sin\theta,0,\cos\theta)$), where $\theta$ is the tilt angle.
Then, we choose an $L_x$ adequately containing the GB transition region and let GB relax in $-L_x \leq x \leq L_x$.
Anchoring boundary conditions are given by the function value and its normal derivatives at $x=\pm L_x$ are equal to the rotated bulk values of two grains (see \cref{sfig:GBSetup} for a schematic), to indicate that $\phi$ is identical to the bulk when $|x|>L_x$. 
The $x$-direction is discretized by the generalized Jacobi polynomials that ensure sufficient accuracy for arbitrary grain orientations. 
The $y$-$z$ plane is discretized by quasiperiodic Fourier expansion \cite{jiang2014numerical,jiang2024numerical}, so that two grains are assumed infinitely large. 
The reciprocal lattice is generated from that of two rotated grains projected on the $y$-$z$ plane. 
Thus, the number of primitive reciprocal vectors varies with the tilt angle. 
When $\sqrt{2} \tan \theta$ is rational (resp. irrational), GBs are $y$-$z$ periodic (resp. quasiperiodic) \cite{friedel1926leccons} and primitive reciprocal vectors are two-dimensional (resp. three-dimensional).
The quasiperiodicity may only occur in the $z$-direction. 
Specifying the primitive reciprocal vectors builds a map from each bulk index to an index in the GB system (see \cref{ssubsec:index} for details). 
The LB energy is minimized in the band region $x\in[-L_x,L_x]$ using the AA-BPG method\,\cite{jiang2020efficient,bao2024convergence}.
We compute the cases for tilt angles throughout $[0,90^\circ]$. 

The spherical structure of FCC can be visualized by a suitable isosurface of $\phi$.
Here we take $\phi=3.5$. 
In bulk FCC, along $[110]$ the spheres are arranged in layers with the spacing $\sqrt{2}a/4$ reappearing every two layers. 
Such an arrangement is maintained in $\left[ 110 \right]$ symmetric tilt GBs, except that some regions enclosed by the isosurface may become nonspherical, which is commonly seen in supramolecular structures and may be regarded as local mobility in atomic crystals. 
Hence, we represent the GB structure by the projection of spheres along the $[110]$ direction. 
To distinguish spheres in two consecutive $(110)$ layers, we color them in blue and white, respectively. 

FCC GBs are characterized by stuctural units formed by a few adjacent spheres, which we illustrate by three periodic GBs with tilt angles $\sqrt{2}\tan\theta=1,2,3$, or with contact planes $(111),(112),(113)$, respectively (\cref{fig:periodic_GBs}), extensively studied in previous works\,\cite{wang2013structure}. 
We denote the structural units as ``$m+n$'' indicating the number of spheres in different $(110)$ layers, where we do not distinguish when blue and white spheres are switched. 
Five structural units are marked in \cref{fig:periodic_GBs}: $A:1+1$; $B:3+1$; $C:2+1$; $D:2+2$ and $E:3+0$.

First, we examine the cases where $ \sqrt{2} \tan \theta$ is a quadratic algebraic number that are associated with quasiperiodic and self-similar generalized Fibonacci sequences\,\cite{horadam1961generalized,decarli1970generalized}. 
\begin{figure*}
	\centering
	\includegraphics[width=1.0\textwidth]{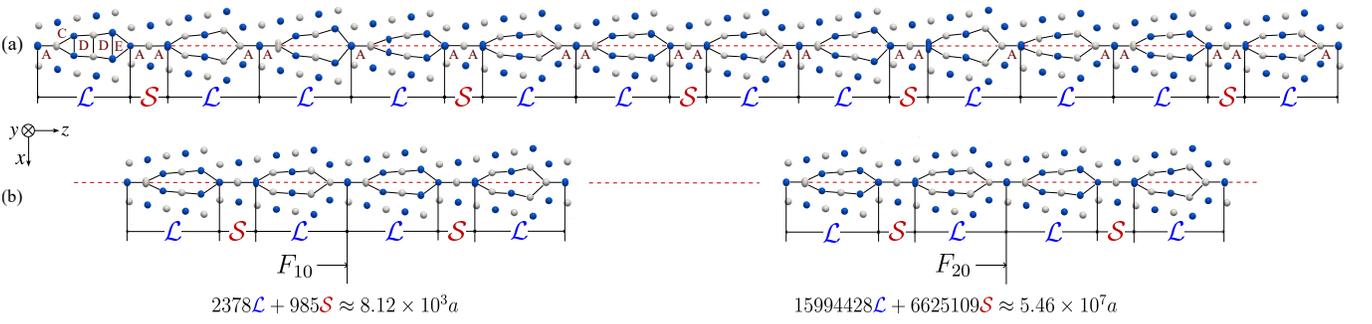}
	\caption{The quasiperiodic GB with tilt angle $\theta = 45^\circ $.
    (a) There are four types of structural units in the GB: $A$, $C$, $D$ and $E$.
    The blue (white) spheres on the GB plane have two kinds of spacing: $\mathcal{L} = (2\sqrt2+3)a/2,~ \mathcal{S} = (\sqrt2+1)a/2$, where $\mathcal{L}/\mathcal{S}=\sqrt 2 +1$.
    (b) $F_{10}$ with $2378 \mathcal{L}$ and $985 \mathcal{S}$, and $F_{20}$ with $15994428 \mathcal{L}$ and $6625109 \mathcal{S}$ are deduced from the substitution rule, showing the GB in regions near their termination positions.}
	\label{fig:45_GB}
\end{figure*}
We present two cases, $\sqrt{2} \tan \theta = \sqrt{2}$ ($\theta=45^\circ$) in \cref{fig:45_GB} and $\sqrt{3}-1$ in \cref{ssec:quasiperiodic_GBs}. 
Structural units $A$, $C$, $D$ and $E$ appear in the GB. 
One $A$, $C$, $E$ and two $D$ form the long spacing $\mathcal{L}= (2\sqrt2+3)a/2$ between two adjacent blue (also white) spheres on the GB plane, while two $A$ form the short spacing $\mathcal{S}= (\sqrt2+1)a/2$.
We infer that the spacing sequence satisfies the substitution rule
\begin{align}
\label{eq:subrule}
	\varrho: \begin{array}{l}
		\mathcal{L} \longmapsto \mathcal{LSL}\\
		\mathcal{S} \longmapsto \mathcal{L}
	\end{array}.
\end{align}
From the legal seed $F_1 = \mathcal{L}$ whose starting position is the left sphere in \cref{fig:45_GB}\,(a), and the definitive mapping $F_{i+1} = \varrho (F_i)$ for $i \geq 1$, we obtain the iterative sequence,
\begin{align*}
    & F_1 \quad \mathcal{L}\\
    \stackrel{\varrho}{\longmapsto} \quad & F_2 \quad \mathcal{LSL} \\ \stackrel{\varrho}{\longmapsto} \quad
	& F_3 \quad \mathcal{LSLLLSL} \\
	\stackrel{\varrho}{\longmapsto} \quad & F_4 \quad
	\mathcal{LSLLLSLLSLLSLLLSL} \\
	\stackrel{\varrho}{\longmapsto} \quad & \cdots
\end{align*}
The sequence satisfies the substitution matrix,
\begin{align}
\label{eq:SubMatrix}
	\begin{pmatrix}
		f^L_{n+1}\\
		f^S_{n+1}
	\end{pmatrix} = \begin{pmatrix}
		2 & 1\\
		1 & 0
	\end{pmatrix} \begin{pmatrix}
		f^L_{n}\\
		f^S_{n}
	\end{pmatrix},~~ n \geq 1, 
\end{align}
where $f^L_n$ and $f^S_n$ are the numbers of $\mathcal{L}$ and $\mathcal{S}$ in $F_n$ and the substitution matrix has the maximum eigenvalue $\mathcal{L}/\mathcal{S}=\sqrt{2} +1$.

Next we compare the two GB sequences derived from substitution rules and numerical computation.
It is easy to find that the first six terms $F_1, \dots, F_6$ of two sequences are the same. Thanks for the used method that can compute the global quasiperiodic GBs, we can obtain the spheres at any given location. 
As an instance, the terms $F_{10}$ and $F_{20}$ obtained by numerical method end at $2378 \mathcal{L} + 985 \mathcal{S} \approx 8.12 \times 10^3 a$ and $15994428 \mathcal{L} + 6625109 \mathcal{S} \approx 5.46\times 10^7 a$, respectively, as shown in \cref{fig:45_GB}(b). We can find that spheres do exist, as predicted by the substitution rule. 
Similar phenomena are also observed at other locations.
As a result, we can conclude that the obtained GB satisfies the quasiperiodic substitution rule \cref{eq:subrule}.
Another quasiperiodic GB ($\theta=\arctan(\sqrt{3}-1)/\sqrt{2}$) that satisfies the generalized Fibonacci substitution rule tied to $\sqrt{3} + 1$, accompanied by some local exchanges of $\mathcal{L}$ and $\mathcal{S}$, is also determined
(see \cref{sfig:273678_GB}).

\begin{figure}[h]
	\centering
	\subfigure[$ \theta = 34^{\circ} $]{
	\includegraphics[width=0.48\textwidth]{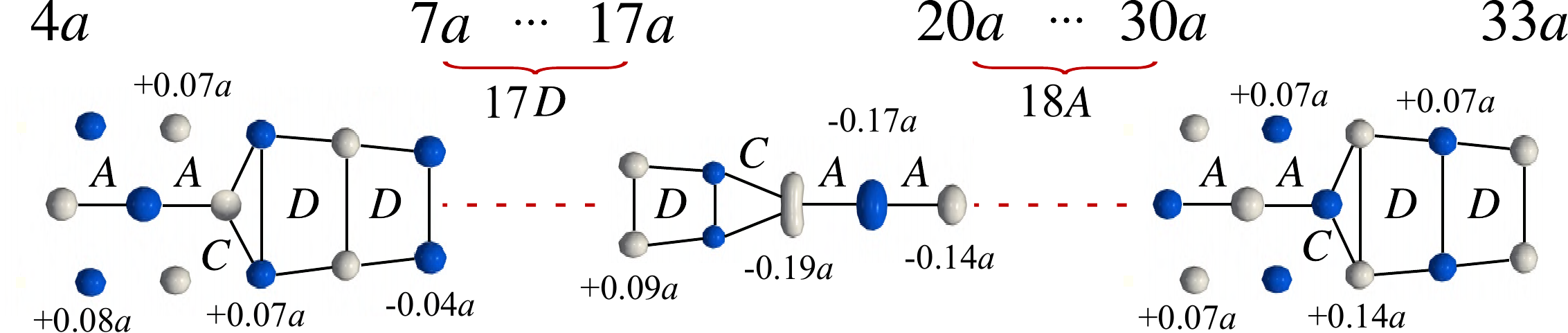}}
    \subfigure[$ \theta = 35^{\circ} $]{
	\includegraphics[width=0.48\textwidth]{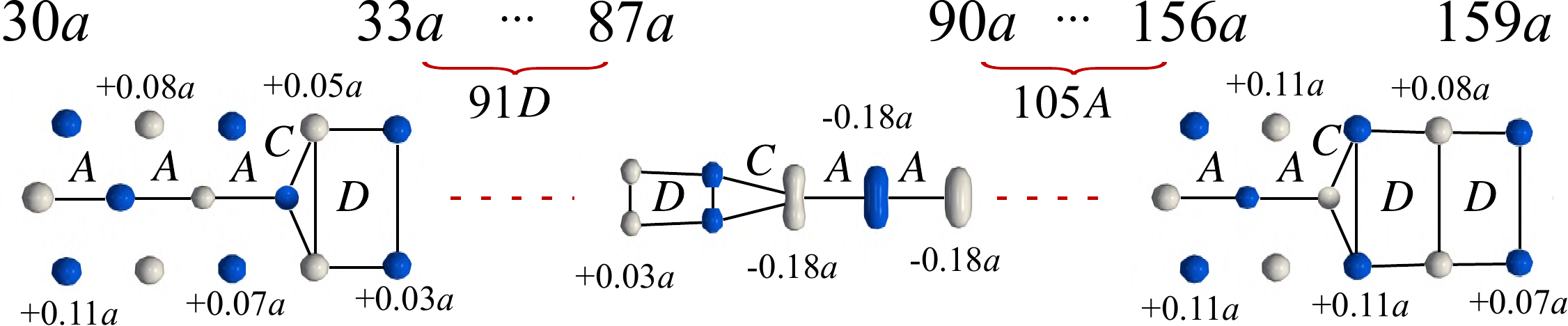}}
    \subfigure[$ \theta = 37^{\circ} $]{
	\includegraphics[width=0.48\textwidth]{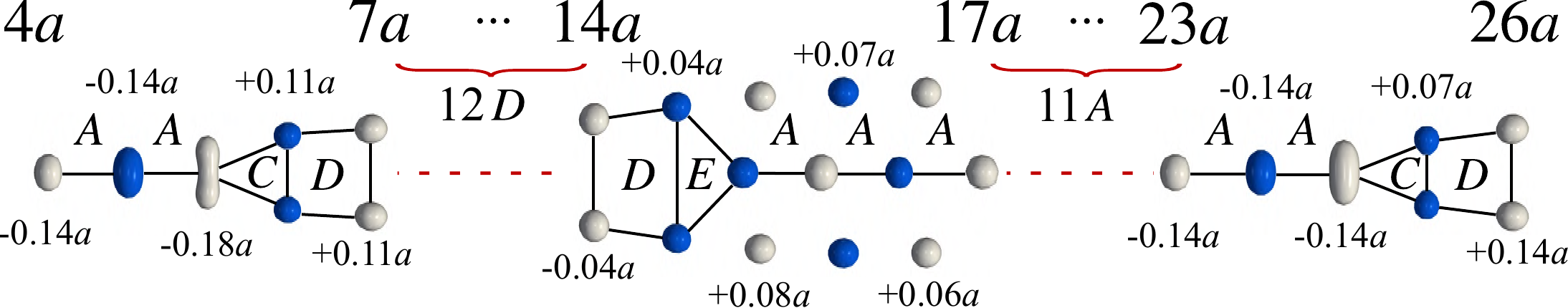}}
	\caption{Quasiperiodic GBs with tilt angles (a) $\theta = 34^\circ$; (b) $\theta = 35^\circ$; (c) $\theta = 37^\circ$. Displacements of spheres from bulk locations are labeled.
	The omitted region comprises a repetitive arrangement of structural units, either $A$ or $D$.}
	\label{fig:quasiperiodic_GBs}
\end{figure}
We turn to exploring the transition mechanism from quasiperiodic to periodic GBs from both structural and spectral perspectives.
We choose $\Sigma 3\,(111) $ and $\Sigma 11\,(113) $ GBs in \cref{fig:periodic_GBs}(a)(b), and here show quasiperiodic GBs with tilt angles close to the twinning angle.
Other quasiperiodic GBs with tilt angles close to $64.76°$ ($\Sigma 11\,(113) $ GB) and the corresponding transition mechanisms are shown in \cref{ssec:transition}.
\begin{figure*}
	\centering
    \subfigure[$ \theta = 34^{\circ} $]{
	\includegraphics[width=0.23\textwidth]{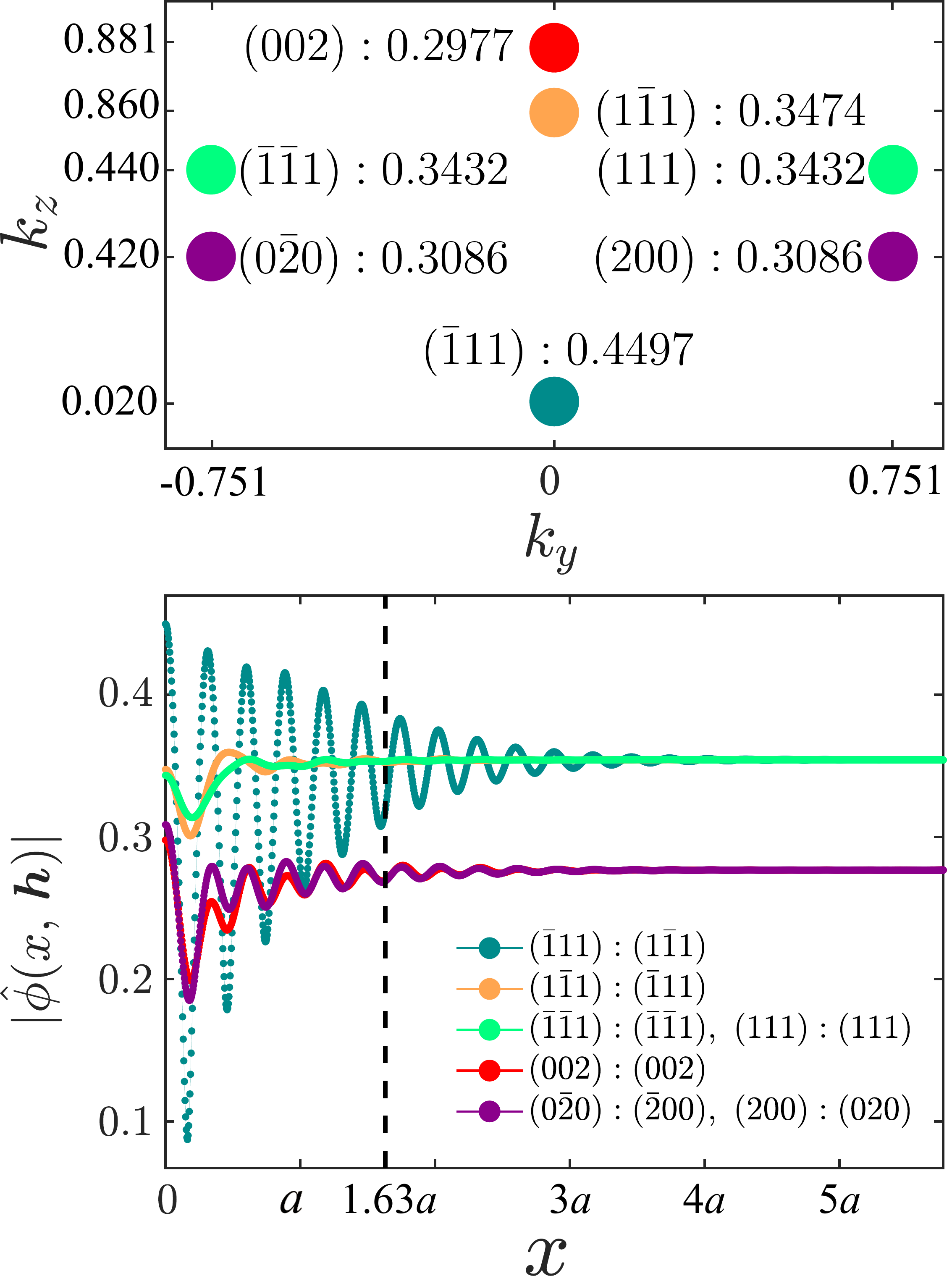}}
	\subfigure[$ \theta = 35^{\circ} $]{
	\includegraphics[width=0.23\textwidth]{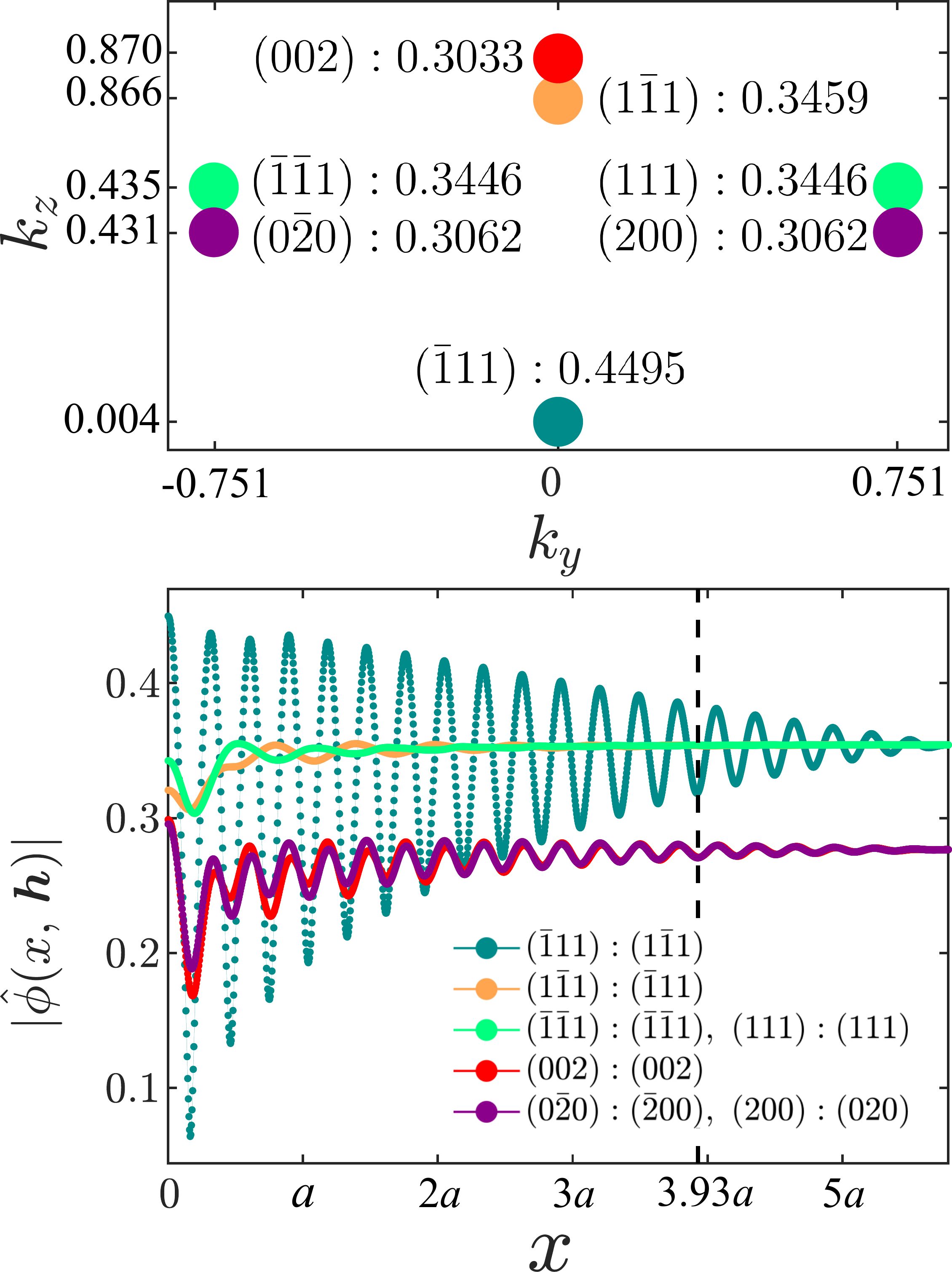}}
	\subfigure[$ \theta \approx 35.26^{\circ} $]{
	\includegraphics[width=0.23\textwidth]{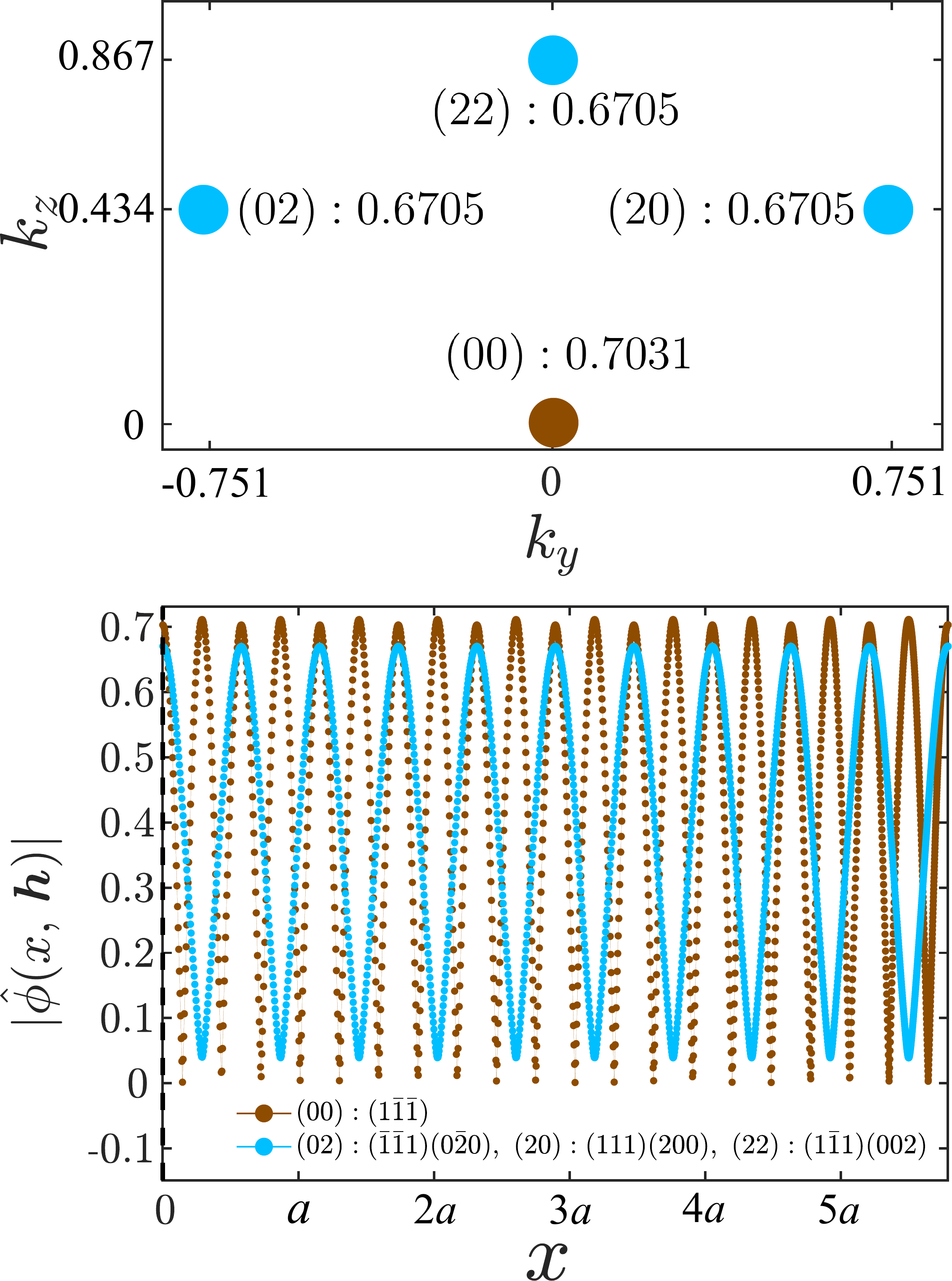}}
	\subfigure[$ \theta = 37^{\circ} $]{
	\includegraphics[width=0.23\textwidth]{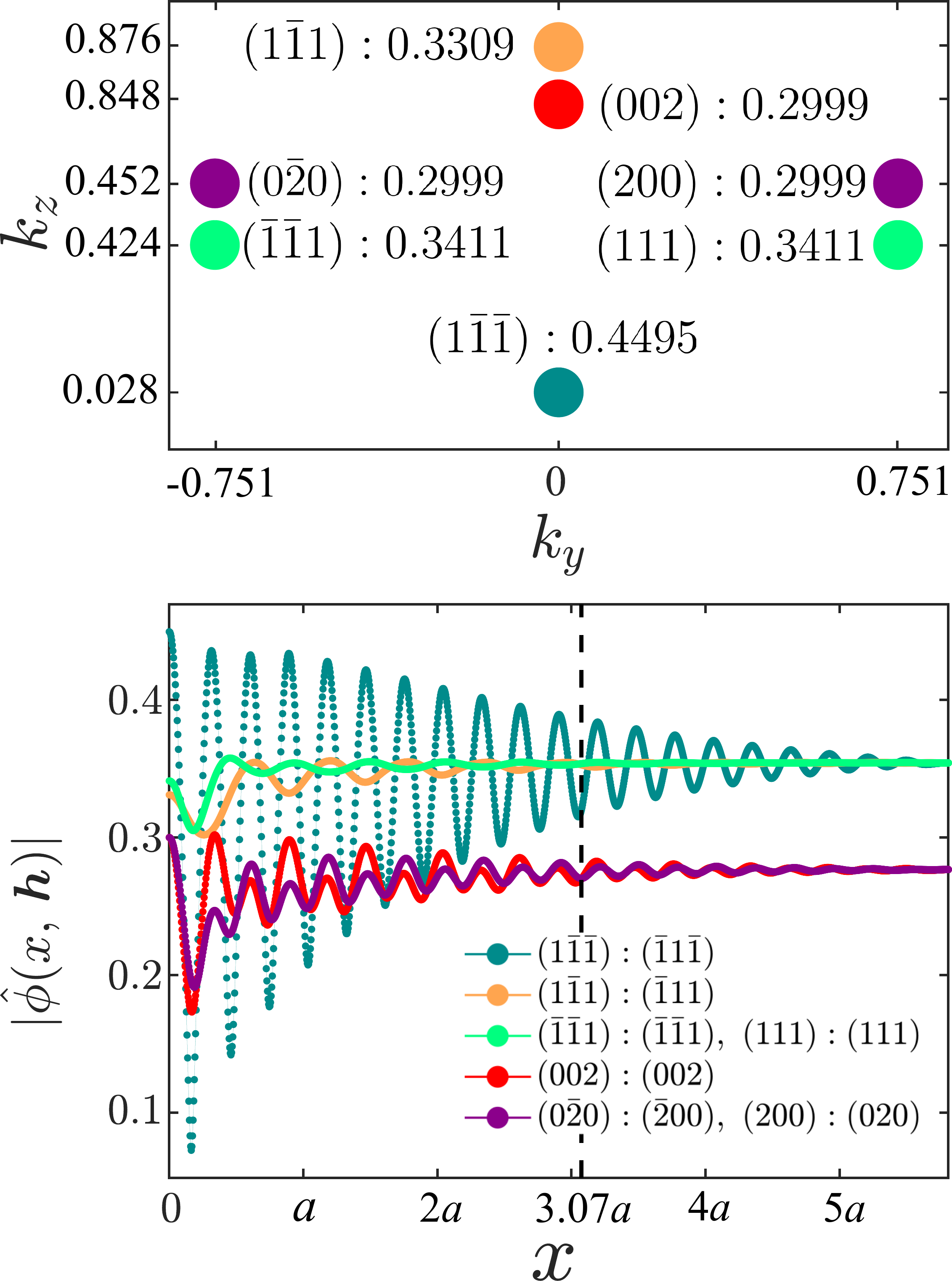}}
	\caption{The distributions of the primary spectra on the $x=0$ plane and their intensities as a function of $x$ with tilt angles of (a) $\theta = 34^\circ$; (b) $\theta = 35^\circ$; (c) $\theta \approx 35.26^\circ$ / $\Sigma 3$ CSL; (d) $\theta = 37^\circ$.
    Relations of spectral indices between GB (before colon) and bulk (after colon) are specified. 
    Only $k_z>0$ and $x>0$ part is presented because of symmetry. Spectra with the same intensity are plotted by a single curve. 
    Black dotted lines represent right edges of GBs defined in \eqref{eq:GB_width} with $ \beta = 0.15$.}
	\label{fig:spectra}
\end{figure*}
\cref{fig:quasiperiodic_GBs} illustrates quasiperiodic GB structures for $\theta = 34^\circ,\, 35^\circ,\, 37^\circ$, near the twin boundary (TB) of $\theta\approx 35.26^\circ$ (\cref{fig:periodic_GBs}(a)). 
Differing from the sole building structural unit $A$ in TB, within quasiperiodic GBs, $A$ and $D$ are arranged in repetition, interrupted by $C$ or $E$.
The closer the tilt angle is to the twinning angle, the more frequently $A$ and $D$ repeat.
The positions of spheres near the GB plane undergo shifts from those in bulk FCC along the $x$-direction, to alleviate localized stresses caused by the misalignment.
The displacements are labeled in \cref{fig:quasiperiodic_GBs} ranging from $\left[-0.19a, +0.14a \right]$, where $+$ indicates movement away from the GB plane and $-$ denotes proximity to the GB plane. 
Structural units $A$ consist of two types of spheres, either newly generated or fused from bulk ones. The former mainly causes positive displacement, while the latter negative.

To further explore the relation between the GBs in \cref{fig:quasiperiodic_GBs} and TB, we examine the spectral components of these GBs in the $y$-$z$ plane. 
Specifically, the Fourier expansion is written as the sum of $\hat{\phi}(x,\boldsymbol{h})\exp \left[ \sqrt{-1}(k_y y+k_z z) \right]$, where $\boldsymbol{h}$ denotes the spectral index and $\boldsymbol{k}=(k_y,k_z)$ represents the actual location of a spectrum. 
We focus on the dominant spectra, i.e. those with $|\hat{\phi}(x,\boldsymbol{h})|$ greater than 0.2, plotting their actual spectral locations (\cref{fig:spectra} upper) and the intensities versus $x$ (\cref{fig:spectra} lower). 
In bulk FCC, the primary spectra, i.e. those with higher intensities, are classes $\left\lbrace 111 \right\rbrace$ and $\left\lbrace 200 \right\rbrace$\,\cite{wagner1963analysis}.
Their projections onto $y$-$z$ plane still dominate in the GBs, with other spectra of the intensities sufficiently smaller (less than one fifth). 
Hence, we focus on the projections of primary bulk spectra in GBs.
Recall that the spectral indices of periodic GBs are two-dimensional, while those are three-dimensional of quasiperiodic GBs.
The map from bulk spectral indices to GB ones is given in \cref{stab:index}. 
We observe that the spectra of quasiperiodic GBs appear in pair near the spectra of TB, and each pair coalesce into a single spectrum with larger intensity when $\theta$ tends to the twinning angle. 
For quasiperiodic GBs, the spectral intensity fluctuates significantly near $x=0$ while gradually approaches the bulk constant value as $x$ increases. 
The fluctuation range is wider when the tilt angle approaches the twinning angle, and eventually no longer decays to constant as the coalescence of spectra occurs on the twinning angle.

We evaluate the GB width by comparing the spectral intensities of GBs with those of bulk FCC $\left| \hat{\phi}_{\text{bulk}} (x,\boldsymbol{h}) \right|$, defined as the interval exceeding a prespecified threshold, 
\begin{align}\label{eq:GB_width} 
    \begin{aligned}
         \Big| \big|\hat{\phi}(x,\boldsymbol{h}) \big|-\big|\hat{\phi}_{\text{bulk}}(x,\boldsymbol{h})\big| \Big| > \beta \max \big| \hat{\phi}_{\text{bulk}}(x,\boldsymbol{h}) \big| .
    \end{aligned}
\end{align}
Here, a unified value $\beta = 0.15$ is chosen for all GBs to make the widths comparable.
The right endpoint of the interval for GBs with tilt angles of $34^{\circ}$, $35^{\circ}$, $35.26^{\circ}$ and $37^{\circ}$ are $x= 1.63a$, $3.93a$, $0$ and $3.07a$, respectively, are shown in \cref{fig:spectra}.

Finally, we examine the GB energy per area for tilt angles throughout $0^\circ$ to $90^\circ$. 
Since the profile $|x|>L_x$ is assumed identical to bulk FCC for the two grains, the excess energy is contributed within $-L_x\le x\le L_x$.
Therefore, the GB energy per area is calculated as 
\begin{align*}
    \gamma = 2L_{x} \left( E_{\text{GB}} - E_{\text{Bulk}} \right),
\end{align*}
where $E_{\text{GB}}$ and $E_{\text{Bulk}}$ stand for energy per volume of GB system and bulk FCC, respectively. 
A sufficiently large $L_x$ is chosen carefully to ensure that $x=\pm L_x$ traverses through lattice points within grains, thereby enabling an accurate computation of energy (see \cref{ssec:energy}).
\begin{figure}[h]
	\centering
	\includegraphics[width=0.48\textwidth]{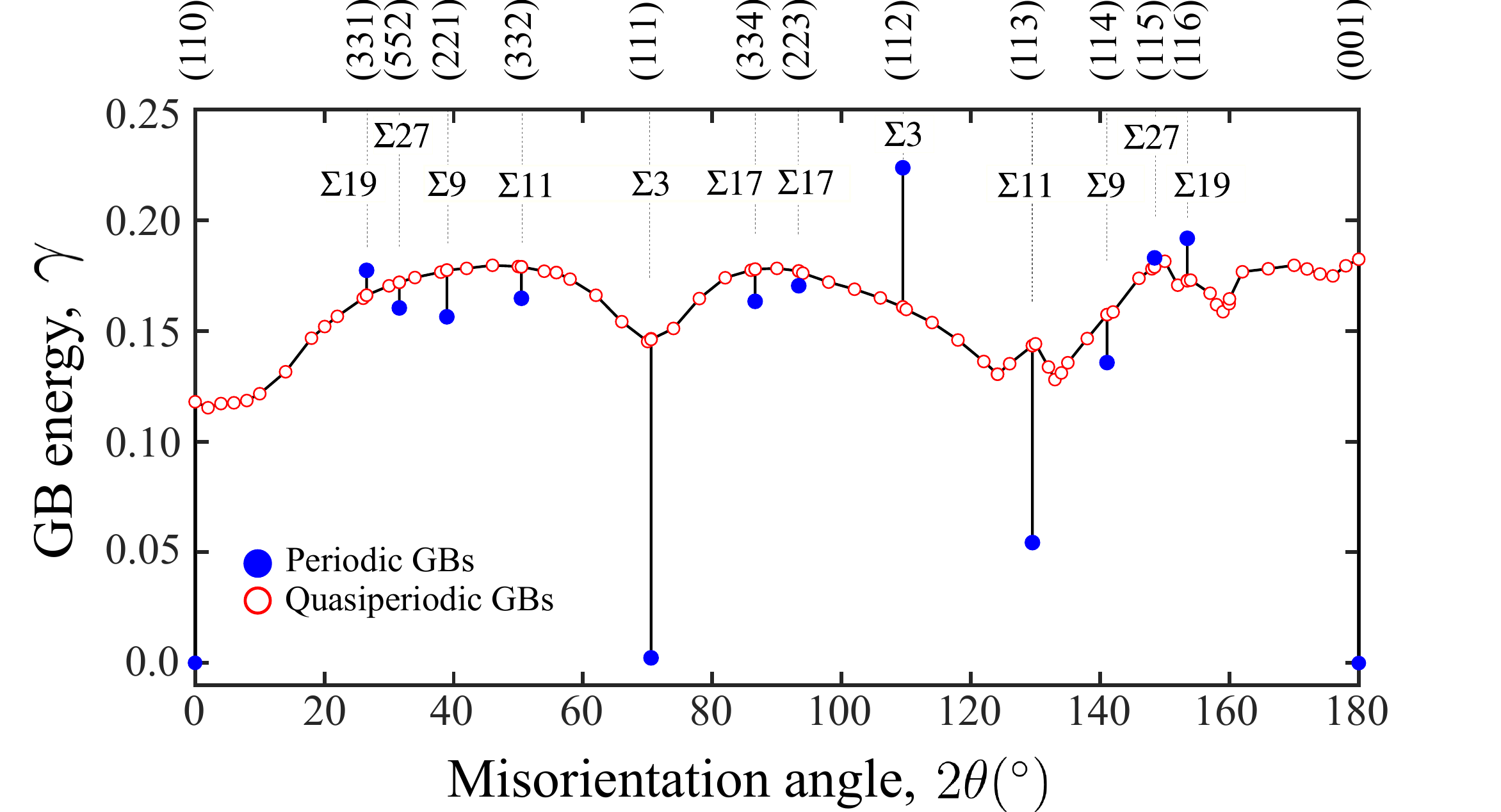}
	\caption{GB energy as a function of  misorientation angle for $\left[ 110 \right]$ symmetric tilt GBs.
	Periodic GBs are represented by blue solid circle, labeled with the $\Sigma$ value and the boundary plane normal direction.
	Quasiperiodic GBs are indicated by red hollow circles.}
	\label{fig:GB_energy}
\end{figure}
The relationship between GB energy and misorientation ($2 \theta$) ranging from $0$ to $180^\circ$ is depicted in \cref{fig:GB_energy}.
The presence of two energy cusps for $\Sigma 3~(111)$ and $\Sigma 11~(113)$ GBs is consistent with existing findings \cite{bulatov2014grain,tschopp2015symmetric}.
It is observed that the GB energy is almost continuous, jumping only at periodic GBs.
The jumps are mostly downward, while a few are upward. 
A considerable amount of spherical rearrangement is observed from quasiperiodic to periodic GBs, despite only minor variations in the tilt angle, potentially accounting for the energy leaps.
The $\Sigma 3~(112)$ GB in \cref{fig:periodic_GBs}(c) is periodic, however, an abundance of vacancies near the boundary leads to its higher energy state.
Compared with previous studies on quasiperiodic GBs, finite-size simulations involve extra surface effects or periodic approximation \,\cite{li2019mechanical,blixt2022grain,jiang2023approximation} that may significant alter the system energy, and our setting accurately prescribes the grain orientations with consistent discretizations.
Real GBs are, although of finite size, generally substantially larger than the finite-size simulations. 
Our results would make it clear the contribution of misorientation apart from other surface effects and periodic approximation error.

In summary, we study $\left[ 110 \right]$ symmetric tilt FCC GBs. 
We discover quasiperiodic GBs of special tilt angles associated with quadratic algebraic numbers that can exhibit generalized Fibonacci sequences. 
The transition mechanism from quasiperiodic GBs to periodic GBs are examined. 
We compute GB energies for general tilt angles, encompassing both periodic and quasiperiodic boundaries, and analyze potential influencing factors on the GB energy.
The methodologies in this work are poised to offer novel perspectives and guidance to the realm of GB investigations.

\begin{acknowledgments}
This work is partially supported by the National K\& D Program of China (2023YFA1008802), NSFC grants (12171412, 12288201, 12371414), and the Innovative Research Group Project of National Natural Science Foundation of Hunan Province of China (2024JJ1008).
We thank the support of High Performance Computing Platform of Xiangtan University.
\end{acknowledgments}

\bibliography{apssamp}

\clearpage
\appendix
\setcounter{subsection}{0}
\setcounter{equation}{0}
\setcounter{figure}{0}
\renewcommand{\thesubsection}{A.\arabic{subsection}}
\renewcommand{\thefigure}{A.\arabic{figure}}
\renewcommand{\thetable}{A.\arabic{table}}
\renewcommand{\theequation}{A.\arabic{equation}}

 \section{Theoretical details\label{ssubsec:TheoDetails}}
 \subsection{Setup of the GB system\label{ssubsec:GBSetup}}
\begin{figure}[h]
 	\centering
 	\includegraphics[width=0.4\textwidth]{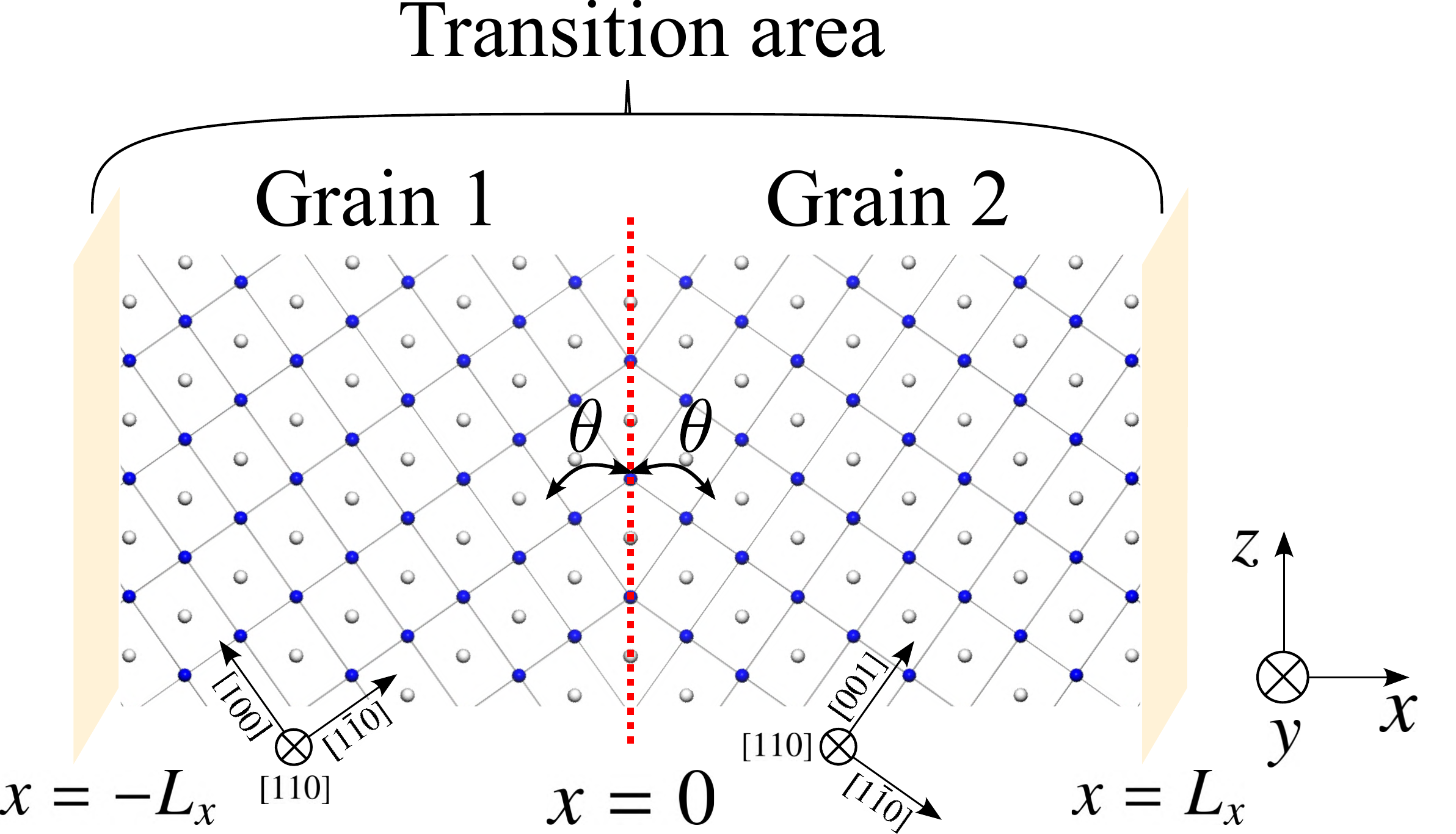}
 	\caption{The schematic of $[110]$ symmetric tilt GBs in FCC.
 	$\theta$ is the tilt angle.}
 	\label{sfig:GBSetup}
 \end{figure}

 The setup of the GB system is shown in \cref{sfig:GBSetup}.
 Before FCC grains are rotated, their $[1\bar{1}0]$, $[110]$, $[001]$ grain directions, respectively, coincide with the $x$,$y$,$z$ axes of the coordinate system.
 Then grain 1 (grain 2) is rotated $\theta$ angle clockwise (counterclockwise) around the $\left[ 110 \right]$ axis.
 The rotated grains occupy two half-spaces, $x<0$ (grain 1) and $x>0$ (grain 2).
 We choose a region $-L_x \leq x \leq L_x$ that contains the entire GB transition area and let GB relax in this region.
 Outside the region the structure is the same as the rotated bulk FCC profile.

 \subsection{Relationship of spectral indices between FCC and GBs\label{ssubsec:index}}
 The order parameter of bulk profile can be expanded by Fourier series, 
 \begin{equation}
 	\phi_{0}(\boldsymbol{r})=\sum_{\boldsymbol{h} \in \mathbb{Z}^{3}} \hat{\phi}_{0}(\boldsymbol{h}) e^{i(\mathcal{P} \boldsymbol{h})^{T} \boldsymbol{r}},
 \end{equation}
 where $\boldsymbol{r}=(x,y,z)^T$ represents the spatial position.
 For FCC, the $3 \times 3$ matrix $\mathcal{P} = bI$ represents the primitive reciprocal lattice of the size $b = 2\pi /a$. 
 After computation, the optimal unit cell length of FCC lattice is $a = 11.8$ when $\tau = 0.2,~ \gamma = 1.5$.
 The integer vector $\boldsymbol{h} \in \mathbb{Z}^{3} $ is the indices of primitive reciprocal vectors.
 Two grains are located in two half-spaces $x<0$ (grain 1) and $x>0$ (grain 2) after different rotations $R_s \in \rm{SO(3)}$, where $s=1, 2$.
 \begin{equation}\label{seq:RotatedGrain1}
 	\begin{split}   
 		\phi_{s}(\boldsymbol{r}) = \phi_{0}\left(R_{s} \boldsymbol{r}\right) 
 		&= \sum_{\boldsymbol{h} \in \mathbb{Z}^{3}} \hat{\phi}_{0}(\boldsymbol{h}) e^{i  \left(R_{sx}^T \mathcal{P} \boldsymbol{h}\right) x } e^{\left(\tilde{R}_{s}^T \mathcal{P} \boldsymbol{h}\right)^{T} \tilde{\boldsymbol{r}}},
 	\end{split}
 \end{equation}
 where $\tilde{\boldsymbol{r}}=(y,z)^T$, $R_s$ is made up of $R_{sx}$ (the first column of $R_s$) and $\tilde{R}_s$ (comprising the second and third columns of $R_s$).
 $\hat{\phi}_{0}(\boldsymbol{h}) e^{i  \left(R_{sx}^T \mathcal{P} \boldsymbol{h}\right) x }$ and $\tilde{R}_{s}^{T} \mathcal{P}$ are rewritten as $\hat{\phi}_{s}(x, \boldsymbol{h})$ and $\tilde{\mathcal{P}}_{s}\in \mathbb{R}^{2 \times 3}$, respectively, and we obtain
 \begin{equation}\label{seq:RotatedGrain2}
 	\begin{split}   
 		\phi_{s}(\boldsymbol{r}) = \sum_{\boldsymbol{h} \in \mathbb{Z}^{3}} \hat{\phi}_{s}(x, \boldsymbol{h}) e^{i\left(\tilde{\mathcal{P}}_{s} \boldsymbol{h}\right)^{T} \tilde{\boldsymbol{r}}}.
 	\end{split}
 \end{equation}
 To construct the least function space that contains two grains, we extract linearly independent column vectors from $(2 \times 6)$-order matrix $(\tilde{\mathcal{P}}_{1},\tilde{\mathcal{P}}_{2})$ to form $(2 \times d)$-order $\tilde{\mathcal{P}}$ such that $ \tilde{\mathcal{P}} \mathbb{Z}^{d} = (\tilde{\mathcal{P}}_{1},\tilde{\mathcal{P}}_{2})\mathbb{Z}^{6}$.
 Grains are reexpressed in the common GB system,
 \begin{equation}
 	\begin{split}   
 		\phi_{s}(x, \tilde{\boldsymbol{r}}) =\sum_{\boldsymbol{h} \in \mathbb{Z}^{d}} \hat{\phi}_{s}(x, \boldsymbol{h}) e^{i\left(\tilde{\mathcal{P}} \boldsymbol{h}\right)^{T} \tilde{\boldsymbol{r}}}.
 	\end{split}
 \end{equation}
 Using primitive reciprocal vectors can establishe the relationship between GB indices and bulk indices.

 For $\left[ 110 \right]$ symmetric tilt GBs with tilt angle $\theta$, rotation matrices of two grains are
 \begin{equation}                                             
 \begin{split}   
 	R_{1} &= \frac{\sqrt{2}}{2} \left(  \begin{array}{ccc}  \cos \theta  & 1 &  \sin \theta \\
 - \cos \theta & 1 & - \sin \theta\\
 -\sqrt{2}\sin \theta & 0 & \sqrt{2} \cos \theta
 \end{array}\right),\\
 R_{2} &= \frac{\sqrt{2}}{2} \left(  \begin{array}{ccc} \cos \theta  & 1 & - \sin \theta \\
 - \cos \theta & 1 &  \sin \theta\\
 ~~\sqrt{2} \sin \theta & 0 & \sqrt{2} \cos \theta
 \end{array}\right).
 	\end{split}
 \end{equation}
 Then we obtain
 \begin{equation}
 	\begin{split}
 		\tilde{\mathcal{P}}_{1} &= \frac{\sqrt{2}b}{2}\left( \begin{array}{ccc}
 			1 & 1 & 0 \\
 			\sin \theta & - \sin \theta  &  \sqrt{2}\cos \theta
 		\end{array} \right),\\
 		\tilde{\mathcal{P}}_{2} &= \frac{\sqrt{2}b}{2} \left( \begin{array}{ccc}
 			1 & 1 & 0 \\
 			- \sin \theta &  \sin \theta & \sqrt{2} \cos \theta
 		\end{array} \right).
 	\end{split}
 \end{equation}
 $\tilde{\mathcal{P}}$ depends on the tilt angle $\theta$.
 When $\sqrt{2} \tan \theta $ is a rational number $p/q~ (p,q \in \mathbb{Z})$, we choose
 \begin{equation}\label{seq:periodic_GB_P}
     \tilde{\mathcal{P}}= \frac{\sqrt{2}b}{2p}\left( \begin{array}{cc}  1 & -1\\ \sin \theta & \sin \theta \end{array} \right),
 \end{equation}
 The index $\boldsymbol{h}$ is two-dimensional, denoted as $(h_1,h_2)^T$.
 GBs in this case are periodic.
 When $\sqrt{2} \tan \theta $ is irrational, we choose
 \begin{equation}
     \tilde{\mathcal{P}}=\frac{\sqrt{2}b}{2}\left( \begin{array}{ccc} 1 & 1 & 0\\ \sin \theta & - \sin \theta  & \sqrt{2}\cos \theta \end{array} \right).
 \end{equation}
 The index $\boldsymbol{h}$ is three-dimensional, denoted as $(h_1,h_2,h_3)^T$.
 GBs in this case are quasiperiodic.

 The spectra in the $y$-$z$ plane are denoted as $\boldsymbol{k} = (k_y, k_z)^T$, represented as $\boldsymbol{k} = \tilde{\mathcal{P}}_2 \boldsymbol{h}$ for grain 2 and $\boldsymbol{k} = \tilde{\mathcal{P}} \boldsymbol{h}$ for GBs.
 The spectra of grain 2, quasiperiodic GBs and periodic GBs are all shown in \cref{stab:index}.
 Taking the index $\boldsymbol{h}=(\bar{1}\bar{1}1)$ as an example, the spectrum of grain 2 is obtained as $k_y = -\sqrt{2}b,~ k_z = 2b/ \sqrt{6} $.
 The twinning angle satisfies $\sqrt{2} \tan \theta = 1$, hence $p$ in \cref{seq:periodic_GB_P} is equal to $1$.
 The spectrum $(k_y, k_z)^T$ is invariant thus TB index satisfies $h_1 - h_2 = -2,~h_1+h_2 = 2$ and yields $h_1=0,h_2=2$.
 The index $(\bar{1}\bar{1}1)$ for grain $2$ corresponds to the index $(02)$ for TB.
 For the other indices, the procedure is similar.
 \begin{table}[h]
     \begin{ruledtabular}
     \begin{tabular}{ccc}
     ~ & $k_y$ & $k_z$ \\
     \hline
     \footnotesize{Grain $2$} & \small{$\frac{\sqrt{2}b}{2} (h_1 + h_2)$} & \small{$\frac{\sqrt{2}b}{2} \left[(h_2 - h_1)\sin \theta + \sqrt{2}h_3 \cos \theta \right]$} \\
     \footnotesize{Quasiperiodic} & \multirow{2}{*}{\small{$\frac{\sqrt{2}b}{2} (h_1 + h_2)$}} & \multirow{2}{*}{\small{$\frac{\sqrt{2}b}{2} \left[(h_1 - h_2)\sin \theta + \sqrt{2}h_3 \cos \theta \right]$}} \\
     \footnotesize{GBs} & & \\
     \footnotesize{Periodic GBs} & \small{$\frac{\sqrt{2}b}{2p} (h_1 - h_2)$} & \small{$\frac{\sqrt{2}b}{2p} (h_1 + h_2) \sin \theta $} \\
     \end{tabular}
     \end{ruledtabular}
     \caption{\label{stab:index}Spectra expressions of grain $2$, quasiperiodic GBs and periodic GBs.}
 \end{table}

 \section{Generalized Fibonacci GBs\label{ssec:quasiperiodic_GBs}}
\begin{figure}[h]
 	\centering
 	\subfigure[$ \theta = \arctan 1 $]{
 		\includegraphics[width=0.23\textwidth]{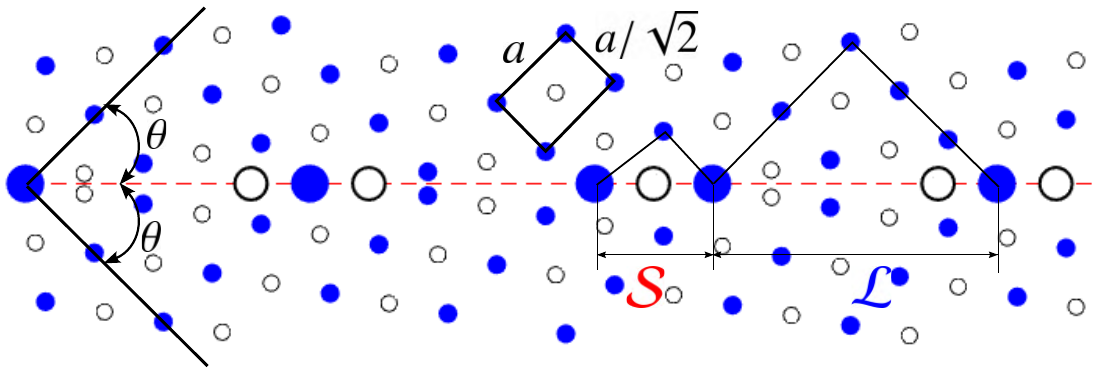}}
 	\subfigure[$ \theta = \arctan \frac{\sqrt{3}-1}{\sqrt{2}}$]{
 		\includegraphics[width=0.23\textwidth]{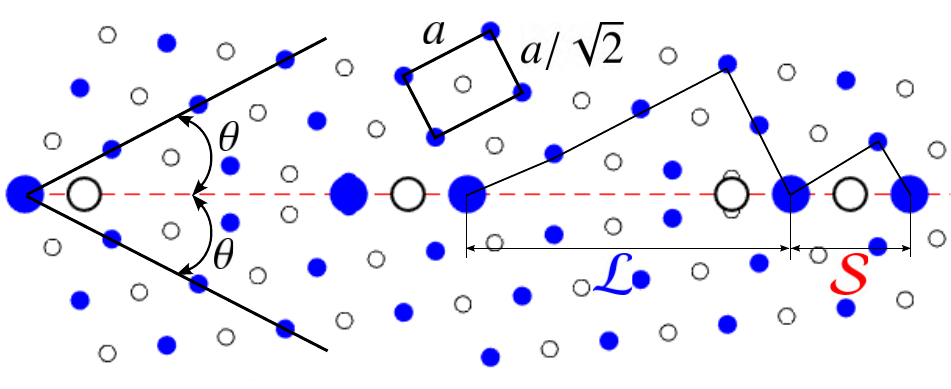}}
 	\caption{The initial configuration of GBs, with tilt angles of (a) $ \theta = \arctan 1 $; (b) $ \theta = \arctan  \left[ (\sqrt{3}-1)/\sqrt{2} \right] $.
 	Two adjacent $(110)$ sphere layers of grains are represented by small blue and white spheres, respectively.
     Spheres of two grains are deflected when they are close enough to the GB plane and merge on the GB plane, denoted by large blue and white spheres.
     Both (a) and (b) have two kinds of spacing $\mathcal{L}$ and $\mathcal{S}$ in the blue and white colored sphere layers.}
 	\label{sfig:non_CSL_grid}
 \end{figure}

 The GB structures depend on grain orientations.
 Quasiperiodic GBs are formed when $\sqrt{2} \tan \theta$ is irrational.
 As shown in \cref{sfig:non_CSL_grid}, when $\sqrt{2} \tan \theta = \sqrt{2}$ and $\sqrt{3} -1$, we place two grains in prespecified orientations, assuming that spheres of the two grains are close enough to offset and fuse to form a sphere on the GB plane, labeled with a large sphere.
 There are two kinds of spacing between two adjacent blue (also white) spheres on GB plane: $\mathcal{L}$ and $\mathcal{S}$.
 They may be quasiperiodic arrangements.

 \begin{figure*}
 	\centering
 	\includegraphics[width=0.75\textwidth]{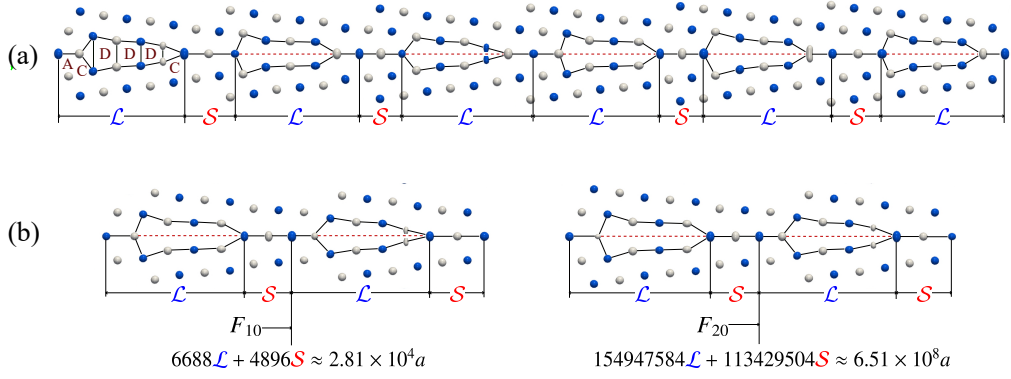}
 	\caption{The quasiperiodic GB with tilt angle $\theta = \arctan ((\sqrt{3}-1)/ \sqrt{2}) $.
     (a) There are three types of structural units in the GB: $A$, $C$ and $D$.
     The blue (white) spheres on the GB plane have two spacing: $\mathcal{L} = (5+3\sqrt{3})a/\sqrt{6+2\sqrt{3}}$, $\mathcal{S} = (2+\sqrt{3})a/\sqrt{6+2\sqrt{3}}$.
     (b) The 10th term of the sequence $F_{10}$ with $6688 \mathcal{L}$ and $4896 \mathcal{S}$, and the 20th term $F_{20}$ with $154947584 \mathcal{L}$ and $113429504 \mathcal{S}$ are deduced from the substitution rule, showing the GB in regions near their termination positions.}
 	\label{sfig:273678_GB}
 \end{figure*}
 The tilt angle $\theta$ of the second case is $\arctan  [(\sqrt{3}-1)/\sqrt{2}]$.
 After computing, we obtain a quasiperiodic GB, as shown in \cref{sfig:273678_GB}.
 The GB have three types of structural units: $A$, $C$, and $D$. 
 The long spacing between two neighboring blue (white) spheres on the GB plane $\mathcal{L}= (3\sqrt2+5)a/\sqrt{6+2\sqrt{3}}$, consisting of one $A$, two $C$ and three $D$.
 The short spacing $\mathcal{S}= (\sqrt3+2)a/\sqrt{6+2\sqrt{3}}$, consisting of two $A$.
 The ratio $\mathcal{L}/\mathcal{S}=\sqrt{3}+1$.
 Elements $\mathcal{L}$ and $\mathcal{S}$ form a sequence that satisfies the substitution rule,
 \begin{align}
 \label{eq:subrule3}
 	\varrho: \begin{array}{l}
 		\mathcal{L} \mapsto \mathcal{LSLS}\\
 		\mathcal{S} \mapsto \mathcal{L}
 	\end{array}.
 \end{align}
 From the legal seed $F_1 = \mathcal{L}$ whose starting position is the left sphere in \cref{sfig:273678_GB}(a), and the definitive mapping $F_{i+1} = \varrho (F_i)$ for $i \geq 1$, we obtain the iterative sequence,
 \begin{align*}
     & F_1 \quad \mathcal{L} \\
     \stackrel{\varrho}{\longmapsto} \quad & F_2 \quad \mathcal{LSLS} \\ \stackrel{\varrho}{\longmapsto} \quad
 	& F_3 \quad \mathcal{LSLSL{\color{red}LS}LSL} \\
 	\stackrel{\varrho}{\longmapsto} \quad & F_4 \quad
 	\mathcal{LSLSLLSLSLLSLS{\color{red}LSLLS}LSLSLLSLS} \\
 	\stackrel{\varrho}{\longmapsto} \quad & \cdots
 \end{align*}
 A few $\mathcal{L}$ and $\mathcal{S}$ interchanges, $\mathcal{LS} \mapsto \mathcal{LSLSL}$ or $\mathcal{LSLLS}$ are acceptable\,\cite{sutton1988irrational}.
 The sequence satisfies the substitution matrix,
 \begin{align}\label{seq:SubMatrix}
 	\begin{pmatrix}
 		f_{n+1}^L\\
 		f_{n+1}^S
 	\end{pmatrix} = \begin{pmatrix}
 		2 & 1\\
 		2 & 0
 	\end{pmatrix} \begin{pmatrix}
 		f_{n}^L\\
 		f_{n}^S
 	\end{pmatrix}, n \geq 1,
 \end{align}
 where $f^L_n$ and $f^S_n$ are the numbers of $\mathcal{L}$ and $\mathcal{S}$ in $F_n$ and the substitution matrix has the maximum eigenvalue $\mathcal{L}/\mathcal{S}=\sqrt{3} +1$.

 We further verify that the obtained GB satisfies the substitution rule \cref{eq:subrule3}. 
 The positions of all spheres are known due to the proposed method that can obtain GBs in the whole space.
 For example, the substitution rule infer that terms $F_{10}$ and $F_{20}$ terminate at $6688 \mathcal{L} + 4896 \mathcal{S} \approx 2.81 \times 10^4 a$ and $154947584 \mathcal{L} + 113429504 \mathcal{S} \approx 6.51\times 10^8 a$.
 Numerical results demonstrate that spheres do exist at these two positions, as shown in \cref{sfig:273678_GB}(b). Similar phenomena can be found at other positions.

 \section{Transition from quasiperiodic GBs to $\Sigma 11 ~(113)$ GB\label{ssec:transition}}
\begin{figure}[t]
 	\centering
 	\subfigure[$ \theta = 63^{\circ} $]{
 		\includegraphics[width=0.48\textwidth]{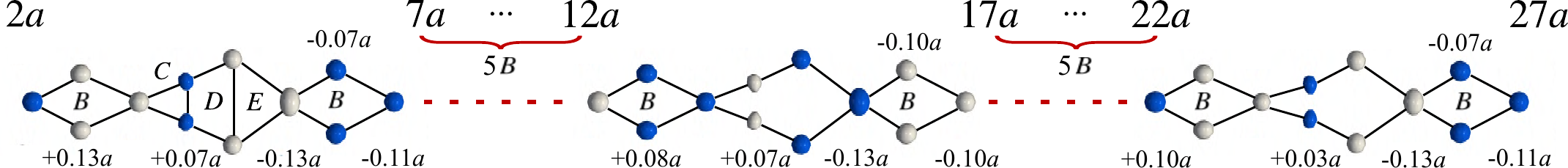}}
 	\subfigure[$ \theta = 64.5^{\circ} $]{
 		\includegraphics[width=0.48\textwidth]{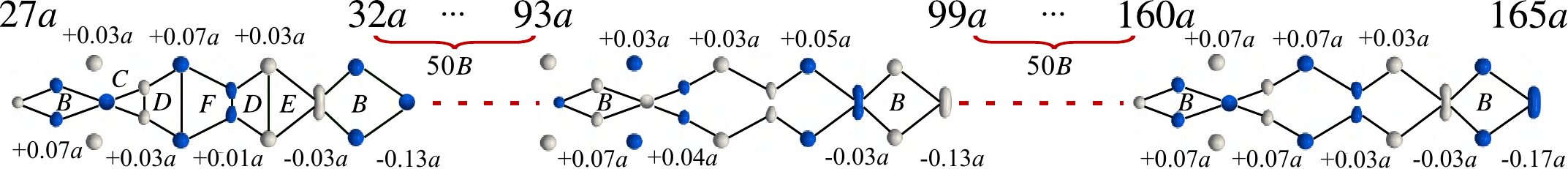}}
 	\subfigure[$ \theta = 66^{\circ} $]{
 		\includegraphics[width=0.48\textwidth]{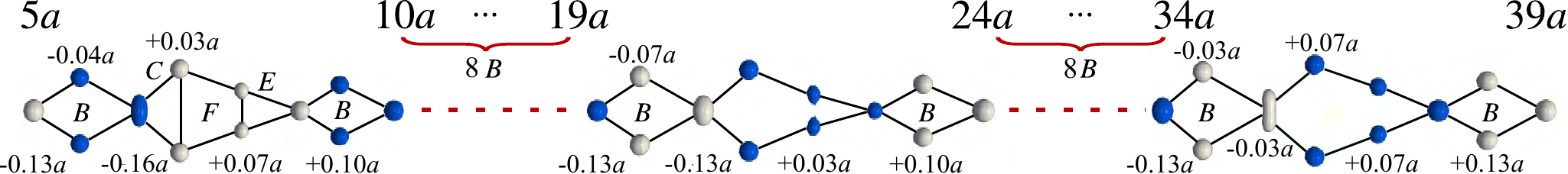}}
 	\caption{Quasiperiodic GBs with tilt angles (a) $\theta = 63^\circ$; (b) $\theta = 64.5^\circ$; (c) $\theta = 66^\circ$.
     Structural units and offsets are labeled.
     The omitted parts are duplicate structural units $B$.}
 	\label{sfig:quasiperiodic_GBs}
 \end{figure}

 \cref{sfig:quasiperiodic_GBs} illustrates quasiperiodic GB structures for $\theta = 63^\circ,~  64.5^\circ,~66^\circ$, near $\Sigma 11~(113)$ GB of $\theta \approx 64.76^\circ$ (\cref{fig:periodic_GBs} (b)).
 Unlike the sole building structural unit $B$ of $\Sigma 11~(113)$ GB, the repeated arrangement of $B$ is interrupted by combinations of $C$, $D$, $E$, and $F:4+0$.
 Their combinations in \cref{sfig:quasiperiodic_GBs}(a)(b)(c) are $CDE$, $CDFDE$, $CFE$, respectively.
 The displacements are labeled in \cref{sfig:quasiperiodic_GBs} ranging from $[-0.17a,~ +0.13a]$.
 Spheres on the GB plane are of two types: newly generated spheres and fusion-formed bulk ones, which cause positive and negative displacement, respectively.
 Both cases result in the structural unit $B$ being duplicated on the GB plane.
 The closer the tilt angle is to the CSL angle $64.76^\circ$, the more repetitions of structural unit $B$ are formed, culminating in $\Sigma 11~(113)$ GB.

 \begin{figure*}[h]
 	\centering
 	\subfigure[$ \theta = 63^{\circ} $]{
 		\includegraphics[width=0.23\textwidth]{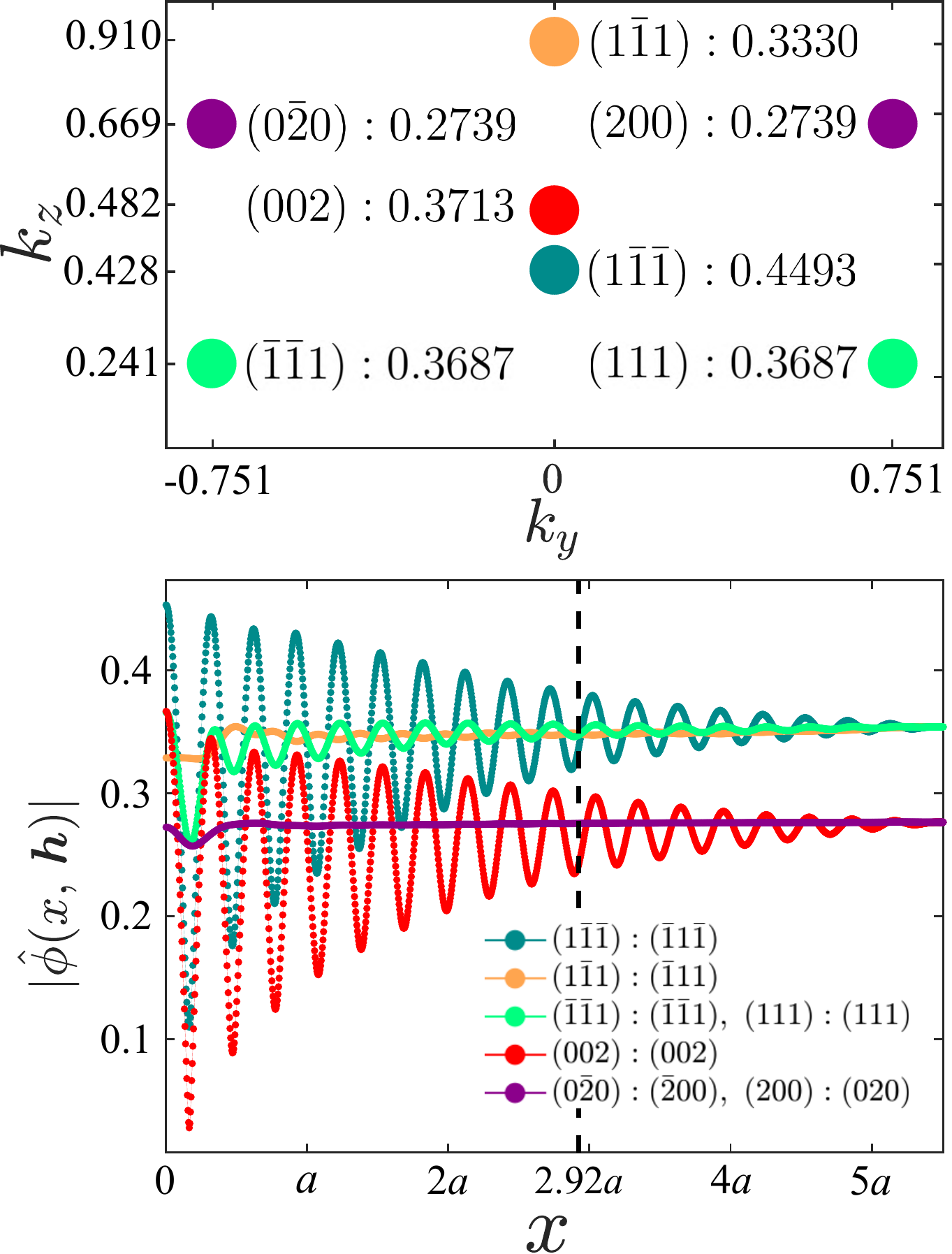}}
 		\subfigure[$ \theta = 64.5^{\circ} $]{
 		\includegraphics[width=0.23\textwidth]{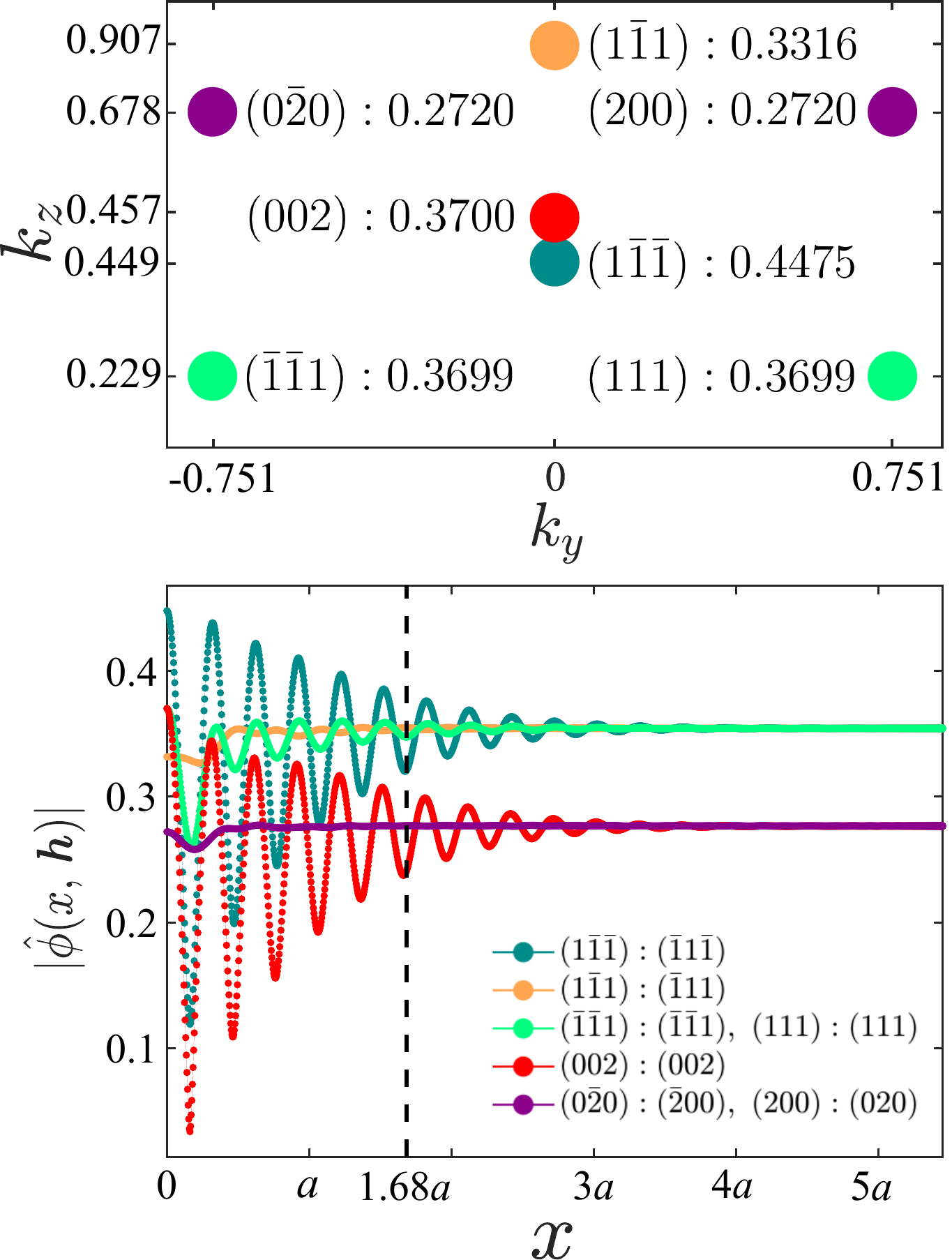}}
 		\subfigure[$ \theta \approx 64.76^{\circ} $]{
 		\includegraphics[width=0.23\textwidth]{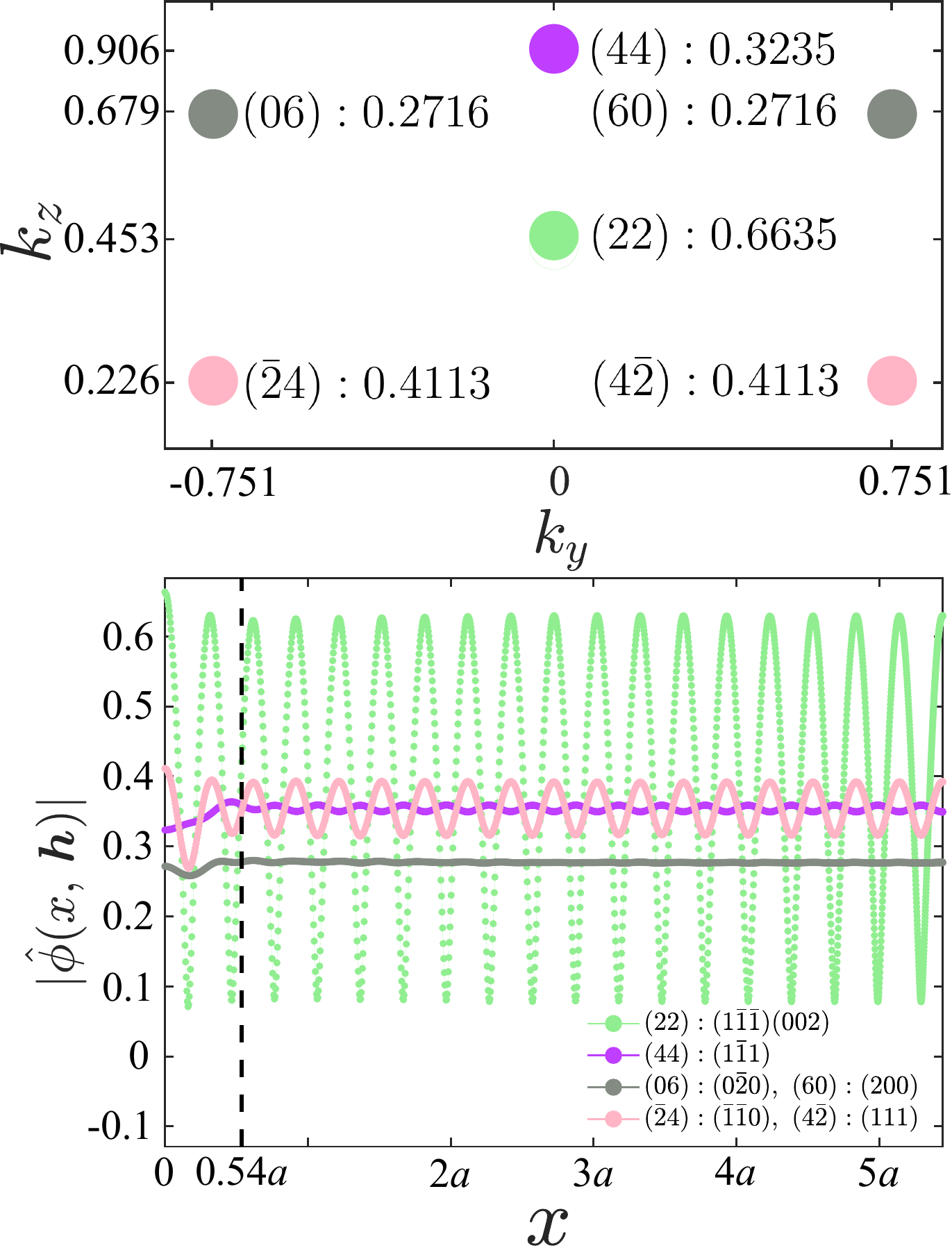}}
 		\subfigure[$ \theta = 66^{\circ} $]{
 		\includegraphics[width=0.23\textwidth]{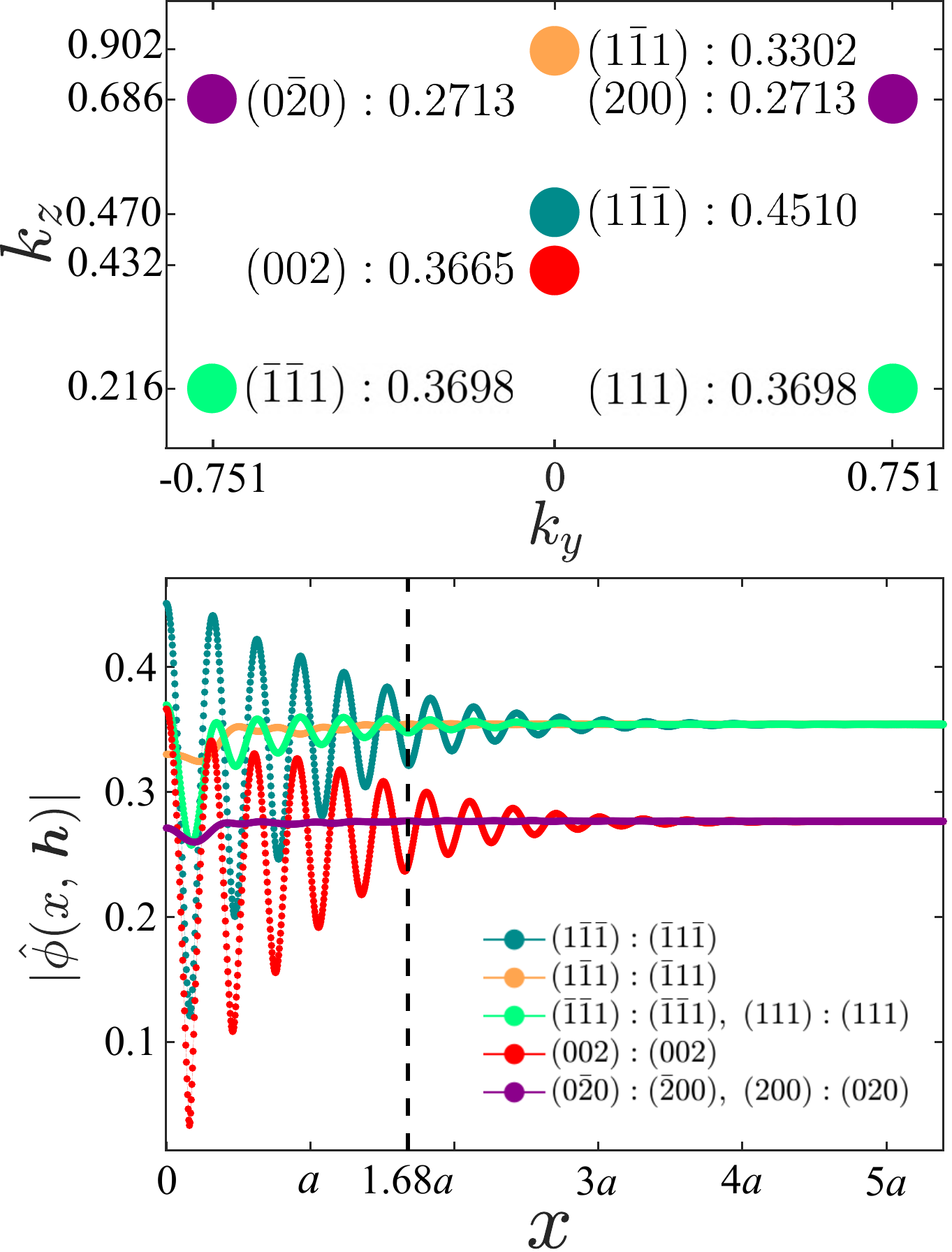}}
 	    \caption{Intensities of primary modes  of GBs with different $ \theta $ against $ x $: (a) $ \theta = 63^{\circ} $; (b) $ \theta = 64.5^{\circ} $; (c) $ \theta \approx 64.76^{\circ} $ / $\Sigma 11$ CSL; (d)  $ \theta = 66^{\circ} $.
 		Black dotted lines represent right edges of GBs defined in text, \cref{eq:GB_width} with $ \beta = 0.15 $.}
 	\label{sfig:spectra}
 \end{figure*}
 To further explore the relation between the GBs in \cref{sfig:quasiperiodic_GBs} and $\Sigma 11~(113)$ GB, we also examine the spectral components of these GBs in the $y$-$z$ plane.
 We focus on the projections of primary bulk spectra in GBs, plot their actual spectral locations (\cref{sfig:spectra} upper) and the intensities versus $x$ (\cref{sfig:spectra} lower). 
 Recall that for periodic GBs the spectral indices are two-dimensional, while for quasiperiodic GBs are three-dimensional. 
 The map from bulk spectral indices to GB spectral indices and is given in \cref{stab:index}.
 $\Sigma 11~(113)$ GBs have more primary spectra compared with TB in \cref{fig:spectra}\,(c), implying that the higher the symmetry the sparser the distribution of primary spectra of CSL GBs.
 The pairs of spectra $(002)(1\bar{1}\bar{1})$ merge into a single spectrum $(22)$ when the tilt angle tends to the CSL angle.
 The intensities of GB spectra exhibit significant fluctuations near $x=0$ while gradually approaches the bulk constant value as $x$ increases.

 Choosing $\beta =0.15$ in \cref{eq:GB_width}, we obtain that the right endpoint of the interval for GBs with tilt angles of $63^{\circ}$, $64.5^{\circ}$, $64.76^{\circ}$ and $66^{\circ}$ are $x=2.92a$, $1.68a$, $0.54a$ and $1.68a$, respectively.

 \section{Method of computing GB energy\label{ssec:energy}}

\begin{figure*}[h]
 	\centering
 		\includegraphics[width=0.4\textwidth]{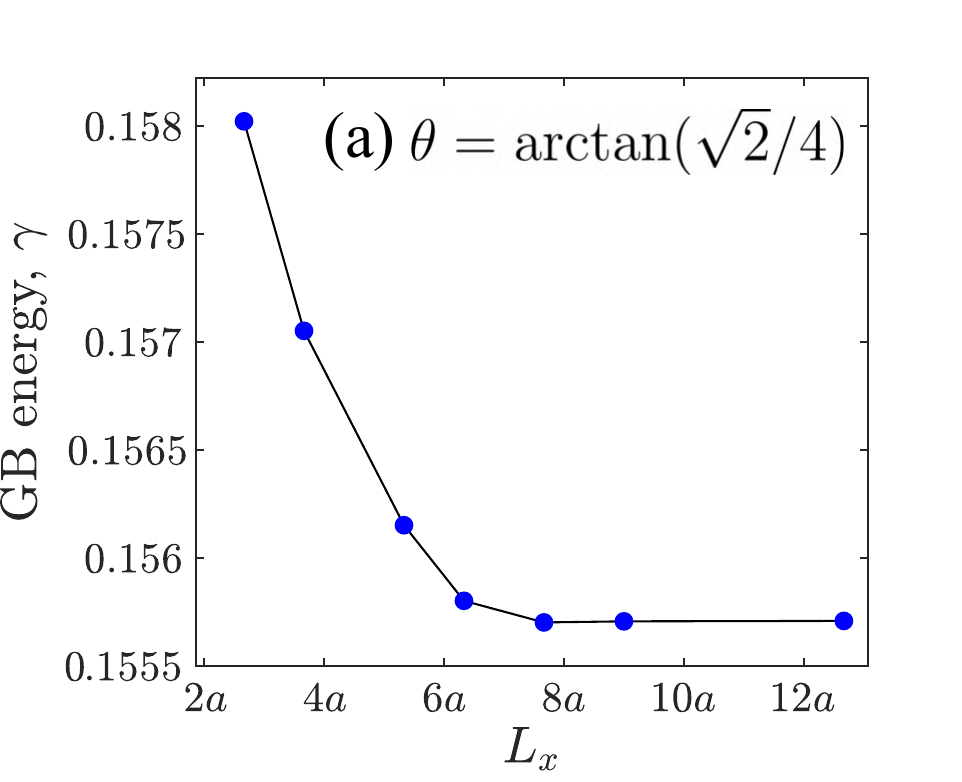}
 		\includegraphics[width=0.4\textwidth]{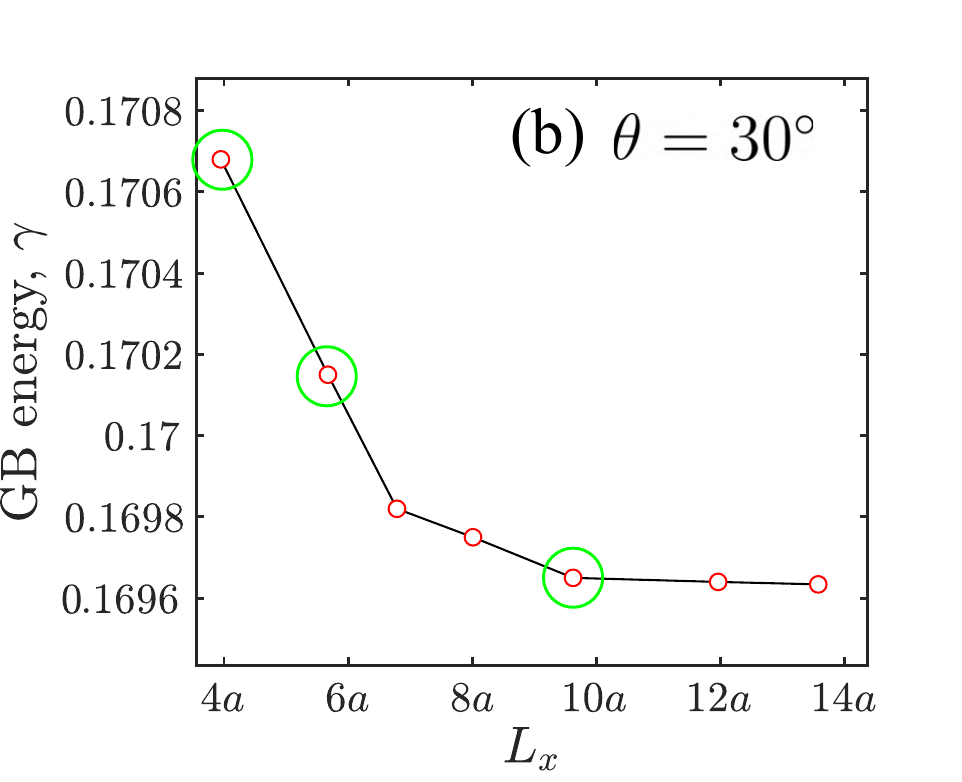}
 	\caption{Two examples of GB energies convergence with increasing $ L_x $: (a) $ \Sigma 9~(221) $ GB ($ \theta = \arctan (\sqrt{2}/4) $); (b) quasiperiodic GB ($ \theta = 30^{\circ} $).}
 	\label{sfig:GB_energy_converge}
 \end{figure*}

\begin{figure*}[h]
 	\centering
 	\subfigure[$ L_x = (3 \cos{\theta} + 2\sqrt{2} \sin{\theta})a$]{
 		\includegraphics[width=0.3\textwidth]{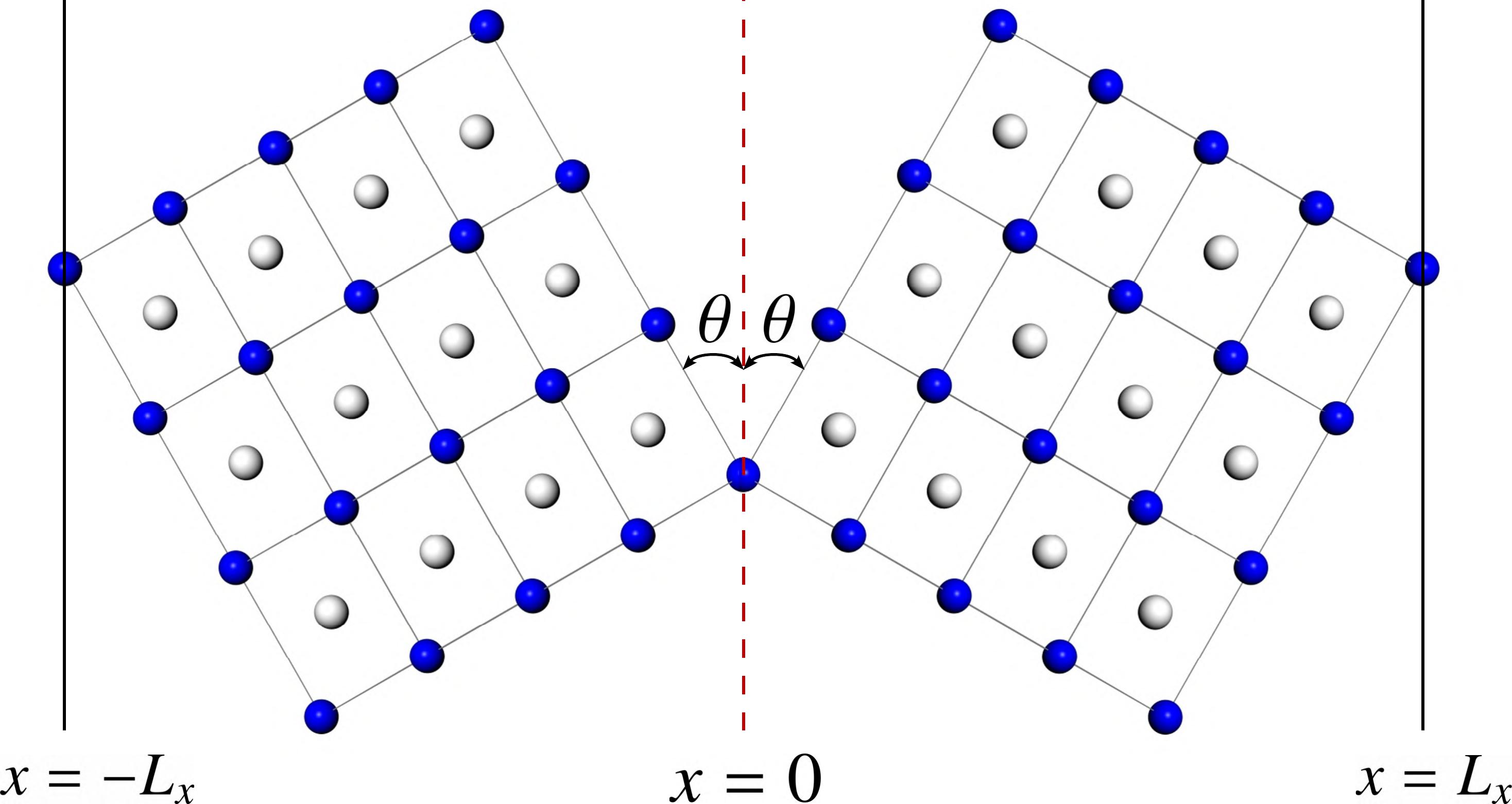}}
 	\subfigure[$ L_x = (4 \cos{\theta} + 3 \sqrt{2} \sin{\theta})a$]{
 		\includegraphics[width=0.3\textwidth]{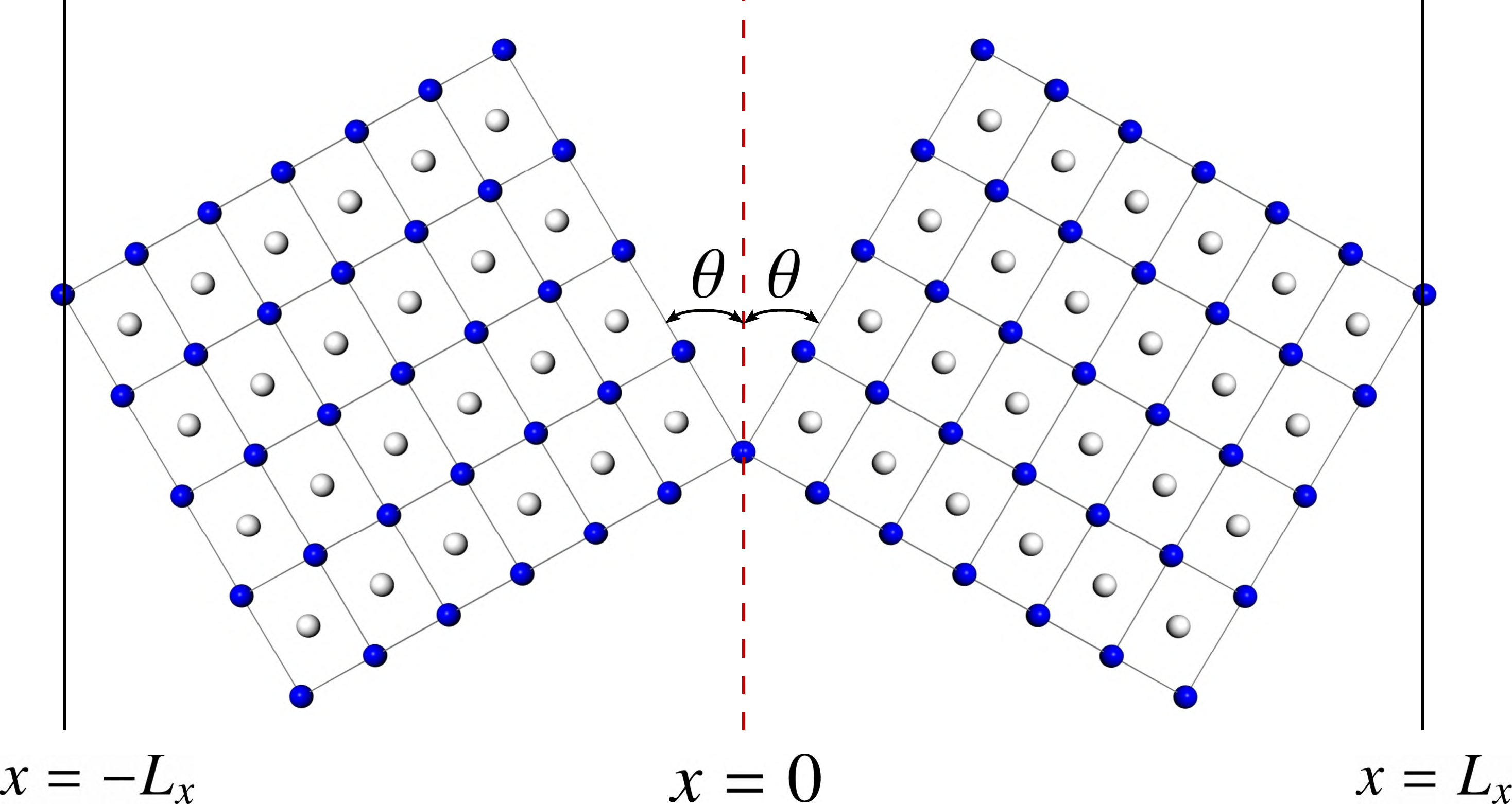}}
 	\subfigure[$ L_x = (7 \cos{\theta} + 5 \sqrt{2} \sin{\theta})a $]{
 		\includegraphics[width=0.3\textwidth]{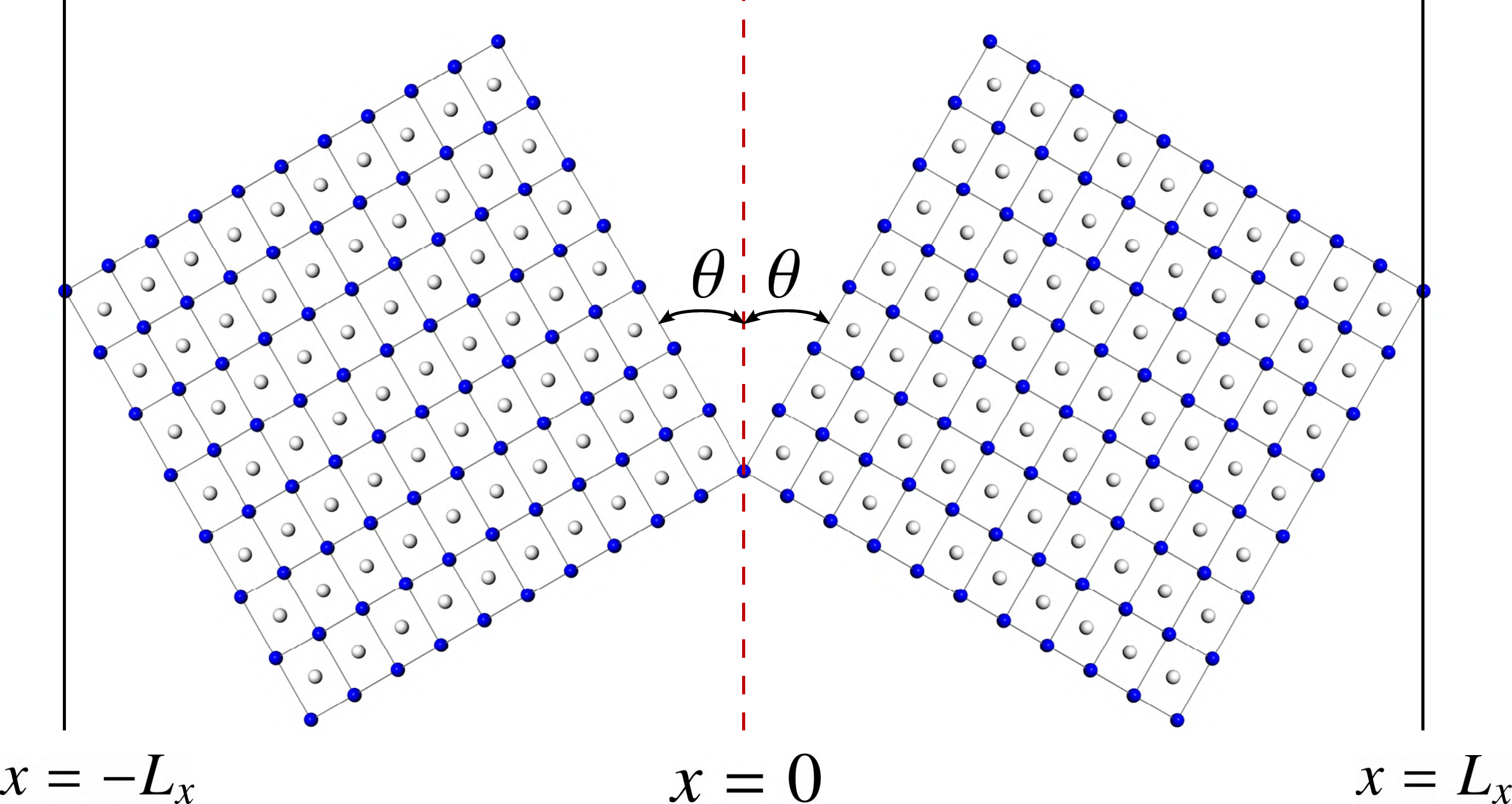}}
 	\caption{Width of transition regions corresponding to three points in \cref{sfig:GB_energy_converge} (b), where $\theta = 30^\circ$.}
 	\label{sfig:Lx}
 \end{figure*}

 We propose an accurate method to calculate GB energy by carefully selecting transition zone. Specially,
 the width $L_x$ of the transition region is chosen to pass through the tilted FCC lattice point for arbitrary tilt angles.
 Here, we present two examples of energy calculations for periodic $\Sigma 9~(221)$ GB ($\theta = \arctan (\sqrt{2}/4)$) and quasiperiodic GB ($\theta = 30^\circ$), respectively, as shown in \cref{sfig:GB_energy_converge}.
 Their GB energies converge with the appropriately increased the width $L_x$, which indicates that GB energies are calculated accurately.
 \cref{sfig:Lx} illustrates three width configurations corresponding to the three green circles in \cref{sfig:GB_energy_converge}(b).
 We opt for a width of $ L_x = (4 \cos{\theta} + 3 \sqrt{2} \sin{\theta})a$ in \cref{sfig:Lx}(b) for computing the GB energy at general tilt angles, resulting in the accuracy of GB energy around $10^{-4}$.

\end{document}